\documentclass[a4paper,10pt]{article}
\pdfoutput=1
\usepackage{amsmath,amsfonts,amssymb}
\usepackage{graphicx}
\usepackage{verbatim}
\usepackage[hyperindex,breaklinks]{hyperref}
\usepackage{url}
\usepackage[width=372pt,height=588pt,top=126pt]{geometry}

\begin{document}

\begin{flushright}
ITP-UU-10/23\\
  SPIN-10/20
\end{flushright}

\begin{center}

\bigskip

\bigskip
\centerline{\LARGE \bf Scalar cosmological perturbations}
\vskip 0.12in
\centerline{\LARGE \bf from inflationary black holes}

\vskip 0.7in

{\Large{\sl Tomislav Prokopec}\footnote{{\it email: }{\tt t.prokopec@uu.nl}} and {\sl Paul Reska}\footnote{{\it email: }{\tt p.m.reska@uu.nl}}}

\vspace{24pt}

Spinoza Institute and Institute for Theoretical Physics\\
Utrecht University\\
Leuvenlaan 4, 3584 CE Utrecht, The Netherlands\\

\end{center}

\vspace{40pt}

\begin{abstract}
\noindent We study the correction to the scale invariant power spectrum of a scalar field on
de~Sitter space from small black holes that formed during a pre-inflationary matter dominated era.
The formation probability of such black holes is estimated from primordial Gaussian density
fluctuations. We determine the correction to the spectrum of scalar cosmological perturbations from
the Keldysh propagator of a massless scalar field on Schwarzschild-de Sitter space. Our results
suggest that the effect is strong enough to be tested -- and possibly even ruled out -- by
observations.
\end{abstract}

\vspace{12pt} \noindent {\small Keywords: primordial black holes, inhomogeneous cosmology, cosmological perturbations, slow-roll inflation}

\date{\today}

\newpage

\section{Introduction}
\label{Introduction}

 Measurements of the cosmic microwave background (CMB) show that the Universe is isotropic on large
scales to very good precision. Indeed, the CMB radiation is an almost perfect black
body~\cite{Mather:1993ij} with a temperature of $T_0=2.725\pm 0.001~{\rm K}$~\cite{Mather:1998gm},
and tiny temperature fluctuations superimposed with an amplitude of the order
$10^{-5}T_0$~\cite{Smoot:1992td}. The exception is the dipole, which is at the level of
$10^{-3}T_0$, and can be explained by our motion with respect to the CMB rest frame. The assumption
that measurements at any position in the Universe would lead to the same result implies that the
Universe is also homogeneous. Observations of large scale structure support this
assumption~\cite{Percival:2009xn}. This Einstein's Cosmological Principle is a corner stone of
modern cosmology and, hence, it is important to further experimentally test its validity.

 There has been much interest recently in the possibility that small violations of homogeneity and/or
isotropy could give rise to the observed CMB anomalies. Some of the often quoted CMB anomalies
are~\cite{Bennett:2010jb,Copi:2010na}: an anomalously small quadrupole and octupole moments; a
large deviation from the mean in some of the higher multipoles; the north-south asymmetry, the
peculiar alignment of the quadrupole and octupole and their pointing in the direction of
Virgo~\cite{de OliveiraCosta:2003pu}; the alignment of some of the higher
multipoles~\cite{Samal:2007nw}; the curious lack of power in the temperature angular correlation
function on large angular scales~\cite{Bennett:2010jb}, {\it etc.} Different authors disagree
however in what constitutes significant deviation from homogeneity and isotropy. For example,
Bennett et al.~\cite{Bennett:2010jb} tend to tune down the statistical significance of these
anomalies, and argue that most of them can be attributed to priors. They also argue that, in the
absence of a deep theoretical justification, which would make further tests possible, these
anomalies will most likely remain curiosities. In addition to the CMB anomalies, there are also
anomalies in the large scale structure of the Universe. For example, observational evidence was
reported by Kashlinsky and collaborators~\cite{Kashlinsky:2009dw} for large scale (dark) flows of
galactic clusters which cannot be explained by homogeneous, adiabatic, Gaussian, cosmological
perturbations generated during inflation. Moreover, some
authors~\cite{Uzan:2009mx,Alexander:2007xx} offer an alternative to dark energy by considering the
earth to be located near the center of a large void~\cite{Alexander:2007xx}, or by considering a
randomly distributed collection of voids in the Universe (the Swiss cheese
Universe)~\cite{Leith:2007ay}.

 The question we pose in this paper is whether some of these anomalies can be explained by placing a
small black hole into an inflationary universe.~\footnote{Our study of small black holes can be
quite easily extended to point-like particles such as magnetic monopoles or heavy particles whose
mass is of the order the Planck mass. However, we expect that the effect of these particle-like
objects will in general be much smaller than that of primordial black holes, giving thus a
competitive edge to the study of black holes in inflation.} Since answering this question
rigorously is hard, here we make a first step in addressing it. In order to model cosmological
perturbations we consider quantum fluctuations of a massless (or light) scalar field minimally
coupled to gravity in Schwarzschild-de Sitter (SdS) space, and calculate the corresponding
spectrum. Based on the knowledge of the Mukhanov-Sasaki gauge invariant potential, we then estimate
the spectrum of scalar cosmological perturbations. We make the assumption that the gauge invariant
treatment also applies to the inhomogeneous cosmology at hand, but warn the reader that our
approach should be tested by a rigorous study of cosmological perturbations in inhomogeneous
settings such as inflation endowed with a small black hole. In this paper we ignore tensor
perturbations, since we expect that their amplitude will be, just as in homogeneous cosmologies,
suppressed when compared to that of scalar perturbations. Furthermore, the effect of vector modes
is not taken into account because they are not dynamical for a homogeneous background and hence we
expect them to play a subdominant role in the case of weak breaking of homogeneity that we
consider.

 It is an important question how to relate our results for the primordial spectrum of scalar
cosmological perturbations to the CMB observables. An interesting study in this direction is
Ref.~\cite{Carroll:2008br}, where the authors investigate how different types of violation of
homogeneity and isotropy would affect the temperature fluctuations in the CMB. Based on symmetry
considerations, the authors consider in particular how a point-like defect (particle), a line-like
defect (cosmic string) or a plane-like defect (domain wall) would modify the observed temperature
anisotropies. By symmetry a small black hole is closest to a point-like object, yet its event
horizon makes it a more complex object to study.~\footnote{A further complication is in the fact
that a black hole could rotate and/or move with respect to the inflaton's rest frame. The latter
could in principle be related to the claimed large scale dark flows~\cite{Kashlinsky:2009dw}.}
While the analysis in~\cite{Carroll:2008br} is useful to make a connection between cosmological
perturbations generated in inflation and temperature fluctuations, it is not general enough to suit
our needs. In particular, it cannot be used for a primordial black hole whose comoving distance
from us is small in comparison to the wavelength of the perturbation considered.

 The main theoretical motivation for studying spectral inhomogeneities generated by a stationary
black hole in inflation is that they yield results that can be tested against observations. This is
so because the resulting spectrum can be viewed as a six parameter {\it template}. A good analogue
are the gravitational wave templates provided by black hole binary systems. To illustrate more
precisely what we mean, recall that homogeneous inflation produces a (power law) spectrum which, as
a function of spatial momentum $k$, can be viewed as a two-parameter template, the parameters being
the spectral amplitude ($\Delta_{\cal R}$) and its slope ($n_s-1$), which have been by now tightly
constrained by CMB measurements~\cite{Komatsu:2010fb}. When viewed as a template, the SdS spectrum
contains four additional parameters. These constitute the black hole position $\vec y\,$ with
respect to us and its mass $M$ during inflation~\footnote{Of course, the black hole has by now
evaporated.}, which we parametrize by $\mu=(GMH_0/2)^{1/3}$. Here $H_0$ denotes the de~Sitter
Hubble parameter, and $G$ is Newton's constant.~\footnote{If in addition the black hole is moving,
three additional parameters are needed to specify its velocity; if it is rotating, three additional
parameters are needed to specify its angular momentum; if it is charged, one more parameter is
needed. We shall not study here observational consequences of these more general settings.
Regarding the results presented in Ref.~\cite{Kashlinsky:2009dw}, it would be of particular
interest to study the spectrum of a (slowly) moving black hole.} In the light of the upcoming CMB
observatories, such as the Planck satellite, and ever increasing large scale galactic redshift
surveys, it is clear that we will be able to test the inflationary black hole hypothesis.

 The paper is organized as follows. In section~2 we present the inhomogeneous
Schwarzschild-de~Sitter background metric and the corresponding scalar field equation of motion. We
make a simple estimate of the number of black holes per Hubble volume at the beginning of inflation
in section 3. Then, in section 4 we derive a formula which relates the amplitude of inflaton
fluctuations on Schwarzschild-de~Sitter space to the spectrum of scalar cosmological perturbations.
The Schwinger-Keldysh propagator is derived in section~5 by expanding in the parameter $\mu$.
Section 6 deals with the application to cosmology. In particular, we obtain the power spectrum and
illustrate its features in various plots in section~7. We close our paper in section~8 with a
discussion. Various technical details are relegated to four appendices. In this paper we work in
units where
%the reduced Planck constant and speed of light in free space
%are equal to unity,
$\hbar=1=c$, but we keep Newton's constant $G=6.674\times 10^{-11}~{\rm m^3 kg^{-1}s^{-2}}$, the
reduced Planck mass, $M_P=(8\pi G)^{-1/2}=2.4\times 10^{18}~{\rm GeV}$ and the Planck mass
$m_p=G^{-1/2}=1.2\times 10^{19}~{\rm GeV}$.

\section{Inflaton field on Schwarzschild-de~Sitter space}
\label{BackgroundSection}

\subsection{Background metric and equation of motion}
\begin{figure}
\centering
\includegraphics[scale=0.9]{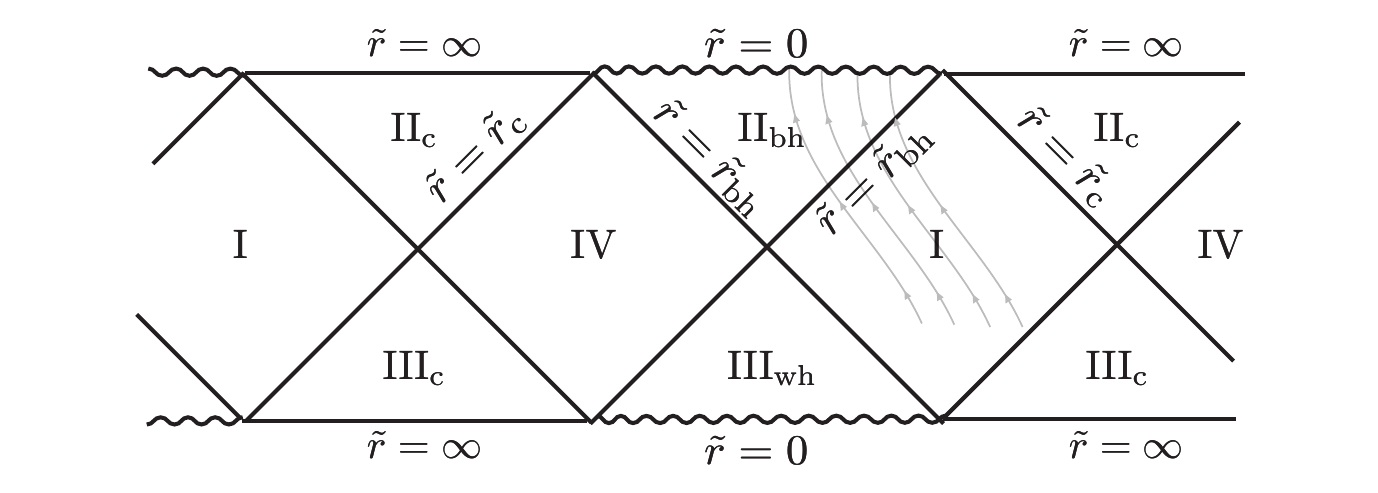}
\caption{\footnotesize This plot shows the Carter-Penrose diagram of Schwarzschild-de~Sitter space.
It is an infinite sequence of black hole regions ($\mathrm{II}_\mathrm{bh}$), white hole regions
($\mathrm{III}_\mathrm{wh}$), static regions ($\mathrm{I}$ and $\mathrm{IV}$) and cosmological
regions ($\mathrm{II}_\mathrm{c}$ and $\mathrm{III}_\mathrm{c}$). The coordinates we use reach the
black hole singularity as well as the cosmological regions. We show schematically some $r={\rm
const.}$ lines (grey). In contrast to $\tilde{r}={\rm const.}$ lines, they are timelike everywhere,
cross the black hole horizon and eventually reach the singularity. Arrows indicate the flow in
conformal time $\eta$. Asymptotic future is given by $\eta\rightarrow 0$ and corresponds to the
boundary $\tilde{r}=\infty$ in the diagram.} \label{fig: PenroseDiagramSdS}
\end{figure}

 A primordial black hole breaks the translational invariance of the background but does preserve
rotational symmetry. The space-time metric of a black hole in an asymptotically homogeneous
universe is the Schwarzschild-de~Sitter (SdS) solution, giving rise to a line element which is
usually written in static coordinates as
\begin{equation}
ds^2 = -f(\tilde{r})dt^2 + \frac{d\tilde{r}^2}{f(\tilde{r})} + \tilde{r}^2d\Omega^2
\,,
\label{staticSdS}
\end{equation}
with $f(\tilde{r})=1-2GM/\tilde{r}-\Lambda \tilde{r}^2/3$. In these coordinates the three
symmetries of the SdS space are manifest: the time translation invariance and the two spatial
rotations. Quantum fluctuations of a scalar field on the Schwarzschild background have been dealt
with in Ref.~\cite{Candelas:1980zt}, where it was found that the radial mode functions of a
massless scalar field can be expressed in terms of Heun's functions. But the presence of a
cosmological horizon complicates the analysis and the SdS case has only been discussed for an
extremal black hole~\cite{Winstanley:2007tf}. The reason for this is the difficult singularity
structure of the d'Alembertian for~(\ref{staticSdS}). For applications to cosmology another form of
the metric is more useful. In Appendix~A we show by explicit coordinate transformations that the
metric takes the form~\footnote{A similar form of the metric can be found in~\cite{McVittie:1933}.
However, it has the disadvantage of being degenerate at the black hole horizon.}
\begin{equation}
ds^2 = a^2(\eta) \Bigg\{-d\eta^2
     + \left(1+\frac{\mu^3\eta^3}{r^3}\right)^{4/3}
           \left[\left(\frac{1-\mu^3\eta^3/r^3}
                            {1+\mu^3\eta^3/r^3}\right)^2dr^2
     + r^2d\Omega^2\right]\Bigg\}
\,,
\label{cosmologicalSdS}
\end{equation}
with $\mu = (GMH_0/2)^{1/3}$ and the scale factor $a$ which is a simple function of conformal time
$\eta$, $a(\eta)=-1/(H_0\eta)$ ($\eta<0$). Notice that the metric~(\ref{cosmologicalSdS}) exhibits
a black hole singularity at a finite radius, $r_0=-\mu\eta$ (see also Eq.~(\ref{Riemann squared})),
such that in this metric $r_0<r<\infty$ covers one half of SdS space. The Carter-Penrose diagram is
plotted in Fig.~\ref{fig: PenroseDiagramSdS} (see also~\cite{Gibbons:1977mu}) which also shows
(schematically) how the interior of the black hole is covered by our coordinates.
 The Hubble rate $H_0$ is related to the potential energy of the inflaton through the Friedmann
equation, $H_0^2=V(\phi_0)/(3M_P^2)$. Here we assume that the inflaton potential energy $V(\phi_0)$
is constant, such that it can be related to the effective cosmological constant as
$\Lambda=V(\phi_0)/M_P^2$.

 The equation of motion for the massless inflaton field $\phi$ is the Klein-Gordon equation,
\begin{equation}
\square \phi(x)=0,
\label{eom:massless scalar in dS}
\end{equation}
where the d'Alembertian $\square$ acting on a scalar field is given by
\begin{equation}
\square \phi(x) = g_{\mu\nu}\nabla^\mu\nabla^\nu \phi(x)
 = \frac{1}{\sqrt{-g}}\partial_\mu \sqrt{-g} g^{\mu\nu}\partial_\nu \phi(x).
\label{scalar dAlembertian}
\end{equation}
One easily finds from the determinant $g$ of the metric tensor,
\begin{equation}
 \sqrt{-g} = a^4r^2\sin\theta\left(1-\frac{\mu^6\eta^6}{r^6}\right)
\,,
\end{equation}
and hence
\begin{align}
\square\phi(x)
 &= \frac{1}{a^2}\Bigg[-\frac{1}{a^2\left(1-\frac{\mu^6\eta^6}{r^6}\right)}
     \partial_\eta a^2\left(1-\frac{\mu^6\eta^6}{r^6}\right)\partial_\eta
\\
 &  + \frac{1}{r^2\left(1-\frac{\mu^6\eta^6}{r^6}\right)}
   \partial_r r^2\frac{\left(1+\frac{\mu^3\eta^3}{r^3}\right)^{2/3}
       \left(1-\frac{\mu^6\eta^6}{r^6}\right)}
            {\left(1-\frac{\mu^3\eta^3}{r^3}\right)^2}\partial_r
   + \frac{1}{r^2\left(1+\frac{\mu^3\eta^3}{r^3}\right)^{4/3}}
           \nabla^2_{S^2}\Bigg]\phi(x)
,
\nonumber
\end{align}
where $\nabla^2_{S^2}$ is the Laplacian on the 2-dimensional sphere,
\begin{equation}
\nabla^2_{S^2}=\frac{1}{\sin(\theta)}
                \frac{\partial}{\partial\theta}\sin(\theta)\frac{\partial}{\partial\theta}
               + \frac{1}{\sin^2(\theta)}\frac{\partial^2}{\partial\phi^2}
\,.
\nonumber
\end{equation}

 Because the CMB is highly isotropic, translation invariance in the early Universe can be only
weakly broken~\cite{Bennett:2010jb,Copi:2010na}. Hence $\mu \ll 1$ and we can expand the metric in
the parameter $\mu$ to first non-trivial order. The result is:
\begin{align}\label{perturbedSdS}
ds^2 &= a^2 \bigg\{-d\eta^2
   + \left(1-\frac{8\mu^3\eta^3}{3r^3}\right)dr^2
   + \left(1+\frac{4\mu^3\eta^3}{3r^3}\right)r^2d\Omega^2\bigg\}
   + \mathcal{O}\left(\mu^6\right)
\\
\sqrt{-g} &=  a^4r^2\sin\theta + \mathcal{O}\left(\mu^6\right)
\,;\qquad
\square = \square^{dS} + \delta\square + \mathcal{O}\left(\mu^6\right)
\,.
\nonumber
\end{align}
The differential operator $\square^{dS}$ is the d'Alembertian on de~Sitter space,
\begin{equation}
\square^{dS}
  = \frac{1}{a^2}\left(-\frac{1}{a^2}\partial_\eta a^2\partial_\eta
                       + \nabla^2\right)
  = \frac{1}{a^2}\left(-\frac{1}{a}\partial_\eta^2 a + \nabla^2
  +\frac{a^{\prime\prime}}{a}\right)
\,,
\label{dAlembertian:dS}
\end{equation}
and
\begin{equation}
\delta \square = \frac{4\mu^3\eta^3}{3r^3}\Bigg(\frac{2r}{a^2}\partial_r
                   \frac{1}{r}\partial_r
           - \frac{1}{a^2r^2}\nabla^2_{S^2} \Bigg)
        = -\frac{4\mu^3H_0^2\eta^5}{3r^3}\Bigg(\nabla^2 - 3\partial_r^2\Bigg)
\,,
\end{equation}
with $a^{\prime} = da/d\eta$ and $\nabla^2=\partial_i\partial_i$ is the Cartesian Laplace operator
in 3 dimensions.

\subsection{An estimate of the perturbation parameter $\mu$}

 The metric~(\ref{cosmologicalSdS}) contains the perturbation parameter $\mu=(GMH_0/2)^{1/3}$ as a
constant. A more realistic point of view, however, is that a black hole of mass $M$ decays due to
an evaporation process and $H_0$ is a time-dependent expansion rate of the Universe. Assuming a
slow-roll inflationary scenario, {\it i.e.}\ small deceleration parameter
\begin{equation}
\epsilon = -\frac{\dot{H}}{H^2} \ll 1
\,,
\end{equation}
we can neglect the time-dependence of $H_0$ in the following discussion. Further comments on this
approximation are made in section~\ref{CMBsection}. For the rate of change of $M$ we make the
following estimate. The evaporation time of a black hole is known to
be~\cite{Carr:1974nx,Saida:2007ru}
\begin{equation}
T = \frac{30720\pi}{g_*}G^2M^3
\,,
\label{EvaporationTime}
\end{equation}
where $g_*\sim 10^2-10^3$ is the number of relativistic degrees of freedom at the energy scale
$\sim 1/(GM)$. The correction to the evaporation time~(\ref{EvaporationTime}) due to the Hubble
horizon is negligible as long as $\mu\ll 1$~\cite{Saida:2007ru}. Assuming that the evaporation
process takes longer than $N=60$ e-foldings, we get
\begin{equation}
\mu > \mu_{\mathrm{crit}}
  = \Bigg(\frac{1}{2}\Big(\frac{N G H_0^2g_*}{30720\pi}\Big)^{1/3}\Bigg)^{1/3}
    \simeq 0.027
\,, \label{mu:limits}
\end{equation}
where we took $G H_0^2\simeq 10^{-12}$ and $g_*\simeq 10^2$. In summary, for primordial black holes
with a mass parameter $0.027\lesssim\mu\ll 1$ we expect the SdS background to be a good realization
of the inhomogeneous inflationary scenario. The constancy of $\mu$ is demonstrated in
Fig.~\ref{fig: Mu_efolding}.
\begin{figure}
\centering
\includegraphics[scale=0.9]{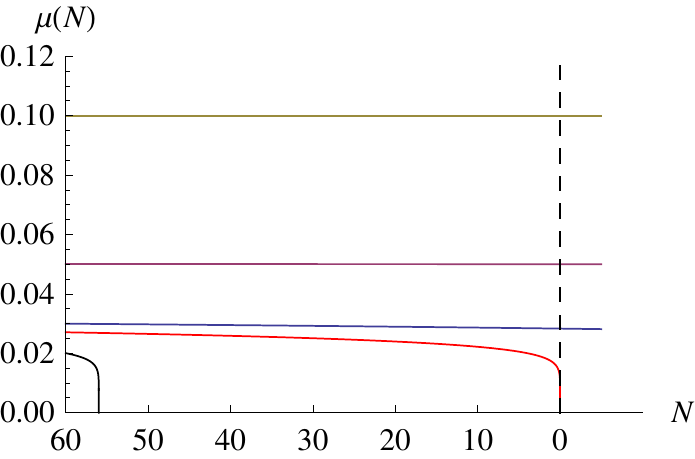}
\caption{\footnotesize The dependence of $\mu$ on the number of e-foldings $N$ is displayed here
for different values $\mu_0$ at the start of inflation. The red curve (second from below)
corresponds to $\mu_\mathrm{crit} \approx 0.027$, the black hole evaporates after 60 e-foldings in
this case. Below this critical value black holes evaporate before the end of inflation, black
(lowest) curve. Above $\mu_\mathrm{crit}$ primordial black holes survive inflation. Their mass is
seen to be constant to a very good approximation (blue, purple and brown curve).} \label{fig:
Mu_efolding}
\end{figure}

\section{Formation probability for black holes}
\label{Formation probability for black holes}

 To assess the physical relevance of primordial black holes, we first estimate the probability for
their formation. For this we consider a pre-inflationary period dominated by heavy non-relativistic
particles with a mass $m<m_p$ and Hubble rate $H_*<m_p$ (here and henceforth an index $*$ refers to
quantities evaluated at the initial time $t_*$). The presence of such particles is generally
expected in physics at the GUT scale~\cite{Georgi:1974sy}. Moreover, such a scenario has recently
been discussed in connection with CMB anomalies~\cite{Fialkov:2009xm,Kovetz:2010kv}. On not too
small scales the matter distribution in this pre-inflationary phase is well described by the local
mass density $\rho(\vec{x},t) = \rho_0(t)(1+\delta(\vec{x},t))$. For black hole formation we are
interested in the growth of density fluctuations inside of a bounded region. The number of
particles in a \textit{comoving} ball with physical radius $R=R(t)$ and volume $V_R$ at time $t$ is
denoted by $N(R,t)$. The initial statistical fluctuations for the particle number are assumed to be
Gaussian, $\delta N(R,t)\sim \sqrt{N(R,t)}$ with variance $\sigma(R,t)=\langle\delta
N(R,t)^2\rangle = \langle N(R,t)\rangle$. The mass density in the ball is linearly related to the
local density,
\begin{equation}
\rho(R,t)=\frac{1}{V_R}\int_{\|\vec x\|<R}d^3\vec x\ \rho(\vec x,t),
\end{equation}
which can also be written as $\rho(R,t) = m N(R,t)/V_R$. Fluctuations in the particle number are
thereby easily translated into density fluctuations.

 Black holes can potentially form from gravitational interaction of such fluctuations
\cite{Carr:1974nx}. Due to the attractive nature of gravity, fluctuations can grow. To study their
evolution, we consider the classical equation of motion for $\delta(\vec{x},t) =
\delta\rho(\vec{x},t)/\rho_0(t)$: \footnote{\textit{cf.} Bonometto (ed.), `Modern Cosmology'
(2002), p. 50.}
\begin{equation} \label{DensityPerturbEvolution}
\ddot{\delta} + 2H\dot{\delta} - c_s^2 \left(\frac{\nabla}{a}\right)^2\delta - 4\pi G\rho_0
(1+w)(1+3w)\delta = 0\,,
\end{equation}
with the speed of sound $c_s^2 = (\partial p_0/\partial \rho_0)_S = w$, where $p_0$ is the
background fluid pressure. We take the equation of state parameter $w=p_0/\rho_0$ to be constant.
The Universe is assumed to be dominated by one species of particles in which case one can neglect
entropy fluctuations, $\delta S=0$, which would otherwise appear on the right hand side of
Eq.~(\ref{DensityPerturbEvolution}). When the gravitational term dominates in the above equation
the density perturbation becomes unstable. The critical scale for the perturbation is given by the
Jeans momentum
\begin{equation}
\left(\frac{k}{a}\right)_J = \frac{\sqrt{4\pi G\rho_0 (1+w)(1+3w)}}{c_s},
\end{equation}
which determines when thermal pressure is in balance with the gravitational force. The Jeans length
$\lambda_J=2\pi/(k/a)_J$ reaches the Hubble scale if $w\approx 1/3$. For $w \approx 0$ the Jeans
length is very small compared to the Hubble radius and small black holes can form. We can solve
Eq.~(\ref{DensityPerturbEvolution}) for $k/a \ll (k/a)_J$ (super-Jeans scale) by making the Ansatz
$\delta \varpropto t^\alpha$ and we find that
\begin{equation}  \label{MomSpaceDensityFluc}
\delta(k,t) = \delta_*(k)\left(\frac{t}{t_*}\right)^{\frac{2(1+3w)}{3(1+w)}} + \overline{\delta}_*(k)\left(\frac{t}{t_*}\right)^{-1}.
\end{equation}
We shall neglect the second mode which is always decaying. In decelerating space-times ($w>-1/3$)
the first mode is growing, whereas in accelerating space-times ($-1\leq w\leq -1/3$) fluctuations
always decay due to the repulsive nature of gravity. The amplification actually increases with
increasing $w$ but, as mentioned before, we are only interested in the case $w \approx 0$ since in
this case $\lambda_J/a\ll R_H$, which also has to be satisfied for the growing solution. The
initial density perturbation is given by
\begin{equation}
\delta_*(k)\equiv \frac{\delta\rho_*(k)}{\rho_*(k)} \simeq
\frac{\delta N_*(k)}{N_*(k)} \sim \frac{1}{\sqrt{N_*(k)}}.
\label{delta N}
\end{equation}
Clearly, we have to look at large fluctuations away from the mean value $\langle N\rangle$ to find
a significant probability of black hole formation. It is convenient to translate momentum space
fluctuations $\delta\rho(k,t)$ to fluctuations in a ball of radius $R$ by writing
\begin{equation}
\frac{\delta N(R,t)}{\langle N(R,t)\rangle}
=\frac{\delta\rho(R,t)}{\rho_0(t)}=\int_0^\infty dk\
\frac{\delta\rho(k,t)}{\rho_0(t)}W(kR)
\,, \label{delta N over N}
\end{equation}
with $\delta\rho (k,t) = \langle \delta\rho(\vec{k},t) \rangle_{\theta,\phi}$ being the angle
averaged momentum space mass density  and
\begin{equation}
W(kR)=\frac{2}{\pi}\frac{\sin(kR)-kR\cos(kR)}{kR}
\end{equation}
is the well-known \textit{window function} for spherically distributed matter in a ball of radius
$R$. From relation~(\ref{delta N over N}) we conclude that $\delta N(R,t)/\langle N(R,t)\rangle$
grows precisely as the momentum space fluctuations~(\ref{MomSpaceDensityFluc}),
\begin{equation}
\frac{\delta N(R,t)}{\langle N(R,t)\rangle}
 = \frac{\delta N_*(R_*)}{\langle N_*(R_*)\rangle}
  \left(\frac{H_*}{H}\right)^{\frac{2(1+3w)}{3(1+w)}}
  \,.
\label{AmplificationOfNumber}
\end{equation}
Next, note that
\begin{equation}
  \langle N(R,t)\rangle = \frac{m_p^2H^2R^3}{2m}
  \,,
\label{average:N}
\end{equation}
which follows from the Friedmann equation, $H^2=(8\pi/3)\rho_0(t)/m_p^2$ and from
\begin{equation}
 \rho_0(t)=\frac{3}{4\pi}\frac{\langle N(R,t)\rangle m}{R^3}
 \,.
 \nonumber
 \end{equation}
Now from Eq.~(\ref{delta N}) and~(\ref{average:N}) we can write,
\begin{equation}
 \sigma_*(R_*) = \langle \delta N_*(R_*)^2\rangle
 = \langle N_*(R_*)\rangle = \frac{m_p^2H_*^2R_*^3}{2m}
 \end{equation}
and hence
\begin{equation}
 \sigma (R,t) \equiv \langle\delta
N^2(R,t)\rangle = \frac{\langle N(R,t)\rangle^2}{\langle
N_*(R_*)\rangle}
\left(\frac{H_*}{H}\right)^{\frac{4(1+3w)}{3(1+w)}}
=\frac{m_p^2H^4R^6}{2mH_*^2R_*^3}\left(\frac{H_*}{H}\right)^{\frac{4(1+3w)}{3(1+w)}}
\,. \label{AmplificationOfVariance}
\end{equation}
Due to the expansion of the Universe, the radius of the ball grows as $R(t)=R_* (a/a_*) =
R_*(H_*/H)^\frac{2}{3(1+w)}$ in~(\ref{AmplificationOfVariance}). In a decelerating universe,
$w>-1/3$, the comoving radius grows slower than the Hubble radius, $R_H=1/H$, such that, if
$R_*<1/H_*$ initially at $t=t_*$, it will remain sub-Hubble at later times. This trend reverses
during inflation, in which $w<-1/3$.

 A black hole with Schwarzschild radius $R_S$ forms if, due to statistical fluctuations, the number
of particles in $V_{R_S}$ becomes sufficiently large, $N(R_S,t)>R_Sm_p^2/(2m)$, $(m_p^2=1/G)$.
Using~(\ref{average:N}) and writing $N(R_S,t) = \langle N(R_S,t)\rangle + \delta N(R_S,t)$ one
obtains the condition
\begin{equation}
\delta N > \delta N_{\rm cr}=\frac{m_p^2R_S}{2m} - \langle
N(R_S,t)\rangle
          = \frac{m_p^2R_S}{2m}[1-(HR_S)^2]
\,. \label{deltaNcr}
\end{equation}
Thus, using that the fluctuations $\delta N$ are Gaussian distributed~\footnote{Recall that, from
the central limit theorem, the Gaussian distribution is the large $N$ limit of the Poisson
distribution.},
\begin{equation}
 P(\delta N) = \frac{1}{\sqrt{2\pi\sigma}}{\rm exp}\Big(-\frac{\delta N^2}{2\sigma}\Big)
\,,
 \nonumber
\end{equation}
with $\sigma = \langle\delta N^2\rangle$, the probability that a black hole forms is found to
be~\footnote{In the limit $\delta N^2 \gg \sigma$ the statistical fluctuations could also obey a
power law behavior, $P_\mathrm{crit} (\delta N) \varpropto (\sigma/\delta N^2)^x$, like in the
theory of critical phenomena. We are not going to consider this possibility here in any detail.
Note, however, that for this type of statistical fluctuations more black holes will form.}
\begin{equation}
P\big(\delta N(R_S,t) > \delta N_{\rm cr}(R_S,t)\big) = \int_{\delta N_{\rm cr}}^\infty d(\delta N)
P(\delta N) \approx \frac{1}{2\sqrt{\pi}}\frac{\exp\Big(-\frac{\delta N_{\rm
cr}^2}{2\sigma(R_S,t)}\Big)}{\delta N_{\rm cr}/\sqrt{2\sigma}} ,\label{probability:BH formation}
\end{equation}
where, making use of Eqs.~(\ref{AmplificationOfVariance}) and~(\ref{deltaNcr}),
\begin{equation}
\frac{\delta N_{\rm cr}}{\sqrt{2\sigma(R,t)}} = \frac{m_p}{2H_*\sqrt{mR_*}}[1-(RH)^2]
=\frac{m_p}{2H\sqrt{mR}}[1-(RH)^2]\Big(\frac{H}{H_*}\Big)^\frac{2+3w}{3(1+w)} \gg 1\,. \label{delta
Ncr:inequality}
\end{equation}
The inequality in~(\ref{delta Ncr:inequality}) is needed to correctly evaluate the
integral~(\ref{probability:BH formation}). Notice also that, when that inequality is met, the
probability for black hole formation is (exponentially) suppressed. Clearly, the inequality is
broken for super-Hubble scales, for which $RH>1$. But at super-Hubble scales we expect suppressed
statistical fluctuations, and hence do not trust our analysis anyway (see the comment further
below). Remarkably, up to the $1-(RH)^2$ term, Eq.~(\ref{delta Ncr:inequality}) is time
independent, which also means that the probability for black hole formation~(\ref{probability:BH
formation}) in a ball of constant comoving radius $R$ is time independent. Hence, the growth of
perturbations~(\ref{AmplificationOfNumber}) precisely compensates the decay in
fluctuations~(\ref{delta N}). This might be telling us something deep about gravity. However, we do
not have a simple explanation for this fact. For later purposes it is useful to rewrite our result
for the probability of black hole formation~(\ref{probability:BH formation}--\ref{delta
Ncr:inequality}) as
\begin{equation}
P(\mu,m,H) = \frac{\mu^{3/2}}{1-16\mu^6}\sqrt{\frac{4mH}{\pi
m_p^2}}\left(\frac{H_*}{H}\right)^{\frac{2+3w}{3(1+w)}}
\exp\Bigg({-\frac{m_p^2(1-16\mu^6)^2}{16mH\mu^3}\left(\frac{H}{H_*}\right)^{\frac{2(2+3w)}{3(1+w)}}}
\Bigg) \,, \label{probability:BH formation:2}
\end{equation}
where we made use of the mass parameter $\mu=(GMH/2)^{1/3}$. Obviously, $\mu < 4^{-1/3}$ must be
satisfied in order for a black hole to be sub-Hubble. The question is then how to convert the
probability~(\ref{probability:BH formation}) into the number of black holes formed before inflation
and during inflation. The analysis presented above is meant to provide a rough estimate of the
number of sub-Hubble black holes formed before inflation, and neither takes a proper account of
causality, nor of nonlinear dynamics of over-densities. Staying within this type of reasoning, we
propose to interpret~(\ref{probability:BH formation:2}) as an estimate for the probability that a
black hole formed by the beginning of inflation in a comoving volume $V_R=(4\pi/3)R^3$. The
expected number of black holes per Hubble volume $(4\pi/3)R_H^3$ at the beginning of inflation is
then
\begin{equation}
\langle N_{\mathrm{BH}}(\mu,m,H)\rangle \approx
\frac{P(\mu,m,H)}{(R_S H)^3} = \frac{P(\mu,m,H)}{64\mu^9} \,.
\label{Black holes per Hubble}
\end{equation}
If the Hubble volume today corresponds to $X$ Hubble volumes at the beginning of inflation, then
there will be about $X\langle N_{\mathrm{BH}}(\mu,m,H)\rangle$ pre-inflationary black holes within
our past lightcone. A more detailed discussion on how to relate the number of black
holes~(\ref{Black holes per Hubble}) to the number of (pre-)inflationary black holes potentially
observable today is given in section~\ref{Discussion and Conclusion}. In Fig.~\ref{fig:
FormationProbability} we show how the expected number of black holes~(\ref{Black holes per Hubble})
depends on the black hole mass parameter $\mu$ for different values of particle mass $m$ and
initial Hubble rate $H_*$. We emphasize, however, that in our analysis we make the assumption that
the statistical fluctuations are normally distributed on all scales. This might not be true, as has
been argued for example in~\cite{Carr:1975qj}, where fluctuations of other types are considered on
super-Hubble scales. Causality can limit the size of these statistical fluctuations. It may be more
realistic to assume that on super-Hubble scales surface fluctuations are dominant, $\delta N
\varpropto \sqrt{S}\sim N^{1/3}$ and $\sigma \approx \langle N\rangle^{2/3}$, thus, suppressing the
formation of black holes that are initially super-Hubble. The actual probability for black hole
formation might therefore be smaller in the region $\mu\sim 4^{-1/3}$ than it is shown in
Fig.~\ref{fig: FormationProbability}. Moreover, the formation probability depends strongly on the
mass $m$ of the heavy particles, and yet we do not know much about it. Based on our current
understanding of particle physics and gravity, it is reasonable to assume that $m$ is limited from
above by the Planck mass $m_p$. If $m\ll m_p$, however, the particles would start behaving
relativistically, which would increase the Jeans length and further suppress, or even prevent, the
formation of black holes.
\begin{figure}
\centering
\includegraphics[scale=0.9]{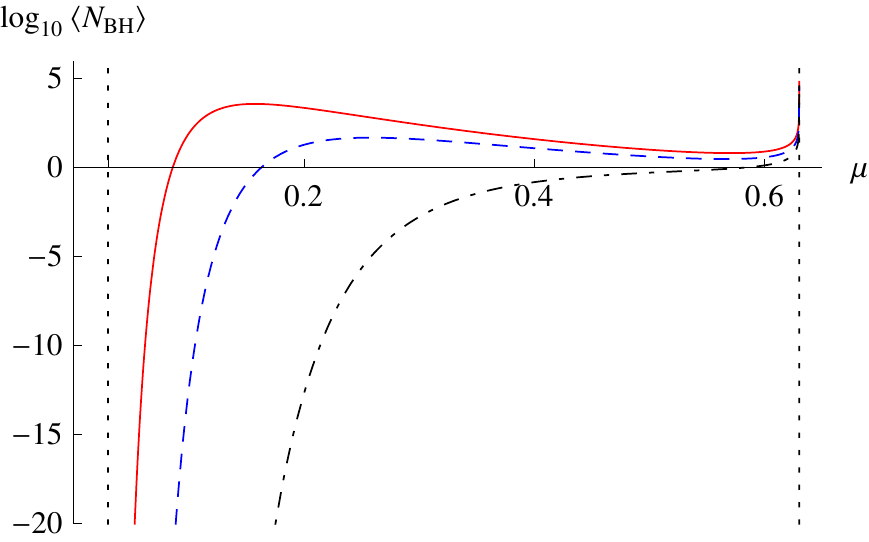}
\caption{\footnotesize In this figure we display the formation probability of black holes per
Hubble volume on a logarithmic scale as a function of the mass parameter $\mu=(GMH/2)^{1/3}$ for
particle mass $m=0.3 m_p$ and initial Hubble parameter $H_*=0.3m_p$ (solid curve, red); $m=0.3 m_p$
and $H_*=0.1m_p$ (dashed, blue) and $m=0.01 m_p$ and $H_*=0.3m_p$ (dot-dashed, black). The left
vertical dashed line indicates the critical black hole mass $\mu_\mathrm{crit} = 0.03$ and the
right one marks $\mu = 4^{-1/3}$. Black holes with mass parameter close to $\mu_\mathrm{crit}$ are
seen to be Gaussian suppressed, whereas in the intermediate range their number can be
$\mathcal{O}(1)$ per Hubble volume at the beginning of inflation.} \label{fig:
FormationProbability}
\end{figure}
\section{From scalar fluctuations to scalar cosmological perturbations}
\label{CMBsection}

\subsection{Homogeneous background}

 Before discussing the effect of inhomogeneities on the scalar spectrum, we review the treatment of
cosmological perturbations on homogeneous backgrounds, such as the conformally flat background
metric, $g_{\mu\nu}^b=a^2(\eta)\eta_{\mu\nu}$, where $\eta_{\mu\nu}={\rm diag}(-1,1,1,1)$ and $a$
is the scale factor of the Universe. The physical situation we have in mind is a slow-roll
inflationary model driven by a homogeneous inflaton field $\phi_0(t)$, where the Hubble parameter
$H(t)=\dot a/a$ ($\dot a=da/dt=da/(ad\eta)$) is a slowly varying function of time, such that the
de~Sitter limit is obtained when $H\rightarrow H_0={\rm const.}$ Scalar perturbations in this model
are induced by the quantum fluctuations of the inflaton, while tensor perturbations are induced by
the quantum fluctuations of the graviton. In linearized perturbation theory the two decouple. For a
recent treatment we refer to~\cite{Prokopec:2010be}, while standard reviews are
Refs.~\cite{Mukhanov:1990me,Kodama:1985bj}.

 It is convenient to decompose the inflaton $\Phi$ and the metric tensor $g_{\mu\nu}$ into the
background fields $\phi_0, g_{\mu\nu}^b$ and the fluctuations $\phi,h_{\mu\nu}$ as follows,
\begin{equation}
\Phi(x) = \phi_0(t) + \phi(x)
\,,\qquad g_{\mu\nu}(x) = g_{\mu\nu}^b(t) + a^2h_{\mu\nu}(x)
\,.
\nonumber
\end{equation}
The background fields $a$ and $\phi_0$ are classical field configurations, whereas $h_{\mu\nu}$ and
$\phi$ are dynamical quantum fields. A detailed study (see {\it e.g.}~\cite{Prokopec:2010be}) shows
that  there are only three physical degrees of freedom, two from the graviton and one from the
scalar field. These can be expressed in terms of the gauge invariant Mukhanov-Sasaki variable
$v=-(\dot{\phi_0}/H){\cal R}$ ($\dot\phi_0=d\phi_0/dt$), and the gauge invariant tensor
$h_{ij}^{TT}$, where by gauge invariance we mean the invariance under linear coordinate shifts
$x^{\mu}\rightarrow x^{\mu}+\xi^\mu(x)$. Since physical observables are independent of the choice
of gauge, it is convenient to work with gauge invariant fields. In homogeneous cosmology the
Sasaki-Mukhanov field (curvature perturbation) is of the form,
\begin{equation}
{\cal R}=\psi - \frac{H}{\dot\phi_0}\phi \,,
\label{SakakiMukhanov}
\end{equation}
where $\psi$ is the scalar gravitational potential defined by the scalar-vector-tensor
decomposition of $h_{ij}$:
\begin{equation}
 h_{ij}= 2\psi\delta_{ij}-2\partial_i\partial_j E
       + 2\partial_{(i}F_{j)} + h_{ij}^{TT}
\label{scalar-vector-tensor}
\end{equation}
and $\phi$ is the inflaton fluctuation. The potential $\psi$ is a gauge variant measure for local
spatial volume fluctuations. When working with gauge invariant variables, such as $\cal R$
in~(\ref{SakakiMukhanov}), we are guaranteed to get observable CMB temperature fluctuations, since
it is the gradient of $\cal R$ that sources photon number fluctuations through the photon Boltzmann
equation. One can also get a physically sensible answer when one fixes a gauge, provided one makes
the correct link to the late time gravitational potential which enters the fluid equations.

 For example, in the comoving gauge, in which $\phi=0=E=F_i=0$, it is the spatial gravitational
potential $\psi$ that determines the Sasaki-Mukhanov field, ${\cal R}=\psi$~(\ref{SakakiMukhanov}).
On the other hand, in the zero curvature gauge, in which $\psi=0=E=F_i=0$, it is the inflaton
fluctuation $\phi$ that determines $\cal R$ through~(\ref{SakakiMukhanov}),
\begin{equation}
{\cal R}= -\frac{H}{\dot{\phi_0}}\phi
  \qquad {\rm (zero\;\; curvature \;\;gauge)}
\,.
\label{SakakiMukhanov:comoving gauge}
\end{equation}
This relation can be used to estimate the late time potential from the inflaton fluctuations
$\phi(x)$. The spectrum of scalar cosmological perturbations ${\cal P}_{{\cal R}}$ in zero
curvature gauge, (${\cal P}_{{\cal R}}$ is conserved on super-Hubble scales) is related to the
spectrum of scalar field fluctuations ${\cal P}_{\phi}$ as,
\begin{equation}
 {\cal P}_{\cal R}
      = \frac{H^2}{\dot{\phi_0}^2} {\cal P}_{\phi}
      = \frac{1}{2\epsilon M_P^2} {\cal P}_{\phi}
\,,
\label{spectrum:comoving gauge}
\end{equation}
where $M_P=1/\sqrt{8\pi G}= 2.43\times 10^{18} \mathrm{GeV} = 4.34 \mu \mathrm{g}$ denotes the
reduced Planck mass. In order to get the latter identity, we used the second Friedmann equation,
$-4\pi G\dot\phi^2 = \dot H$, and the definition for the slow-roll parameter, $\epsilon=-\dot
H/H^2$.

 Of course, there are two remaining degrees of freedom in the graviton which have not been taken
into account. However, the graviton spectrum ${\cal P}_{\rm grav}$ is known to be suppressed in the
slow-roll inflation~\cite{Peiris:2003ff,Komatsu:2010fb,Prokopec:2010be}
\begin{equation} \label{rParameterForGravitonSpectrum}
{\cal P}_{\rm grav}=\frac{2H_0^2}{\pi^2M_P^2}, \qquad
  r\equiv\frac{{\cal P}_{\rm grav}}{{\cal P}_{\cal R}}
     = 16 \epsilon \ll 1
\,,
\end{equation}
such that, to first approximation, we can neglect the graviton contribution to the spectrum of
cosmological perturbations.

 Finally, the field fluctuations can be translated to the temperature-temperature correlation
function as~\cite{Carroll:2008br},
\begin{eqnarray}
\bigg\langle \frac{\delta T (\hat{n}_1)\delta T (\hat{n}_2)}{T_0^2}\bigg\rangle
 &=& \int \frac{d^3k d^3k^\prime}{(2\pi)^6} \sum_{l,l^\prime}
             \frac{(2l+1)(2l^\prime+1)}{(4\pi)^2}
   (-i)^{l+l^\prime}
\label{temperature corr fn}
\\
 &&\hskip 1.5cm
   \times\,  P_l(\vec{k}\cdot \hat{n}_1/k)
        P_{l^\prime}(\vec{k^\prime}\cdot \hat{n}_2/k^\prime)
        {\cal R}(\vec{k})\Theta_l(\vec{k})
     {\cal R}(\vec{k}^\prime)\Theta_{l^\prime}(\vec{k}^\prime)
\,,
\nonumber
\end{eqnarray}
where $P_l(x)$ denotes a Legendre polynomial, and $\Theta_l(\vec k\,)$ and
$\Theta_{l^\prime}(\vec{k}^\prime)$ denote the appropriate transfer functions, which relate ${\cal
R}(\vec k\,)$ to $\delta T(\vec k\,)$, and which are obtained by solving the Boltzmann equation for
the photon fluid.

\subsection{A small black hole in a de~Sitter Universe}

 Keeping in mind the procedure of the previous section for the derivation of cosmological
perturbations from scalar fluctuations, we have to pay special attention to use gauge invariant
fields also in the inhomogeneous case. We split the metric into the Schwarzschild-de~Sitter
background~(\ref{perturbedSdS}) and a perturbation $h_{\mu\nu}$. The spatial part of the
perturbation $h_{ij}$ can be decomposed as
\begin{equation}
 h_{ij}= 2\psi\gamma_{ij}-2\overline{\nabla}_{(i}\overline{\nabla}_{j)} E
       + 2\overline{\nabla}_{(i}F_{j)} + h_{ij}^{TT}\,,
\label{scalar-vector-tensor}
\end{equation}
where $\gamma_{ij}$ is the induced metric on the spatial slices for the metric~(\ref{perturbedSdS})
and $\overline{\nabla}$ is the covariant derivative compatible with $\gamma_{ij}$. The vector
$\xi^\mu$ that generates the gauge transformation $x^\mu \rightarrow x^\mu +\xi^\mu (x)$ can be
written as $\xi^\mu=(\xi^0,\xi^i)$, with $\xi^i = \xi^i_\mathrm{T}+\overline{\nabla}^i \xi$ and
$\overline{\nabla}_i \xi^i_{\mathrm{T}}=0$. The metric perturbation transforms as
\begin{align}
h_{ij} \rightarrow h'_{ij} &= h_{ij} - 2 \nabla_{(i} \xi_{j)} \\
&= 2\psi\gamma_{ij}-2\overline{\nabla}_{(i}\overline{\nabla}_{j)} (E+\xi) +
2\overline{\nabla}_{(i}(F_{j)}-\xi^\mathrm{T}_{j)}) + h_{ij}^{TT} + 2\Gamma^0_{ij}\xi_0\,,
\nonumber
\end{align}
with the Christoffel symbols of the metric~(\ref{perturbedSdS})
\begin{equation}
\Gamma^0_{ij} = \frac{H_0}{a}\gamma_{ij} - \frac{3H_0}{2a}\delta\gamma_{ij}
\end{equation}
and $\gamma_{ij} = \gamma^{\mathrm{dS}}_{ij} + \delta\gamma_{ij} + \mathcal{O}(\mu^6)$ with
\begin{equation}
\gamma^{\mathrm{dS}}_{ij} = a^2 \left(\begin{array}{ccc} 1 & 0 & 0\\ 0 & r^2 & 0\\ 0 & 0 & r^2 \sin^2\theta \end{array} \right), \qquad \delta\gamma_{ij} = \frac{4a^2\mu^3\eta^3}{3r^3}\left(\begin{array}{ccc} -2 & 0 & 0\\ 0 & r^2 & 0\\ 0 & 0 & r^2 \sin^2\theta \end{array} \right).
\end{equation}
Moreover, we observe that $\delta\gamma_{ij}$ is traceless to the relevant order in $\mu$, showing
that $\xi_0$ generates also transformations of $F_j$ and $h_{ij}^{TT}$. The transformation
properties of the two scalars $\psi$ and $E$ are
\begin{equation}
\psi\rightarrow \psi + \frac{H_0}{a}\xi_0 + \mathcal{O}(\mu^6), \qquad E\rightarrow E +\xi + \mathcal{O}(\mu^6).
\end{equation}
Together with the gauge transformation of the scalar field, $\phi \rightarrow \phi +
(\phi_0'/a^2)\xi_0$, we find that
\begin{equation} \label{SasakiMukhanovFieldForSdS}
\psi-\frac{H_0 a}{\phi_0'}\phi \rightarrow \psi-\frac{H_0 a}{\phi_0'}\phi + \mathcal{O}(\mu^6).
\end{equation}
We conclude that the Sasaki-Mukhanov field $\cal R$ in Eq.~(\ref{SakakiMukhanov}) is gauge
invariant (to order $\mu^3$) also when the inhomogeneity caused by a small black hole is taken into
account.

 The fact that $\xi_0$ generates gauge transformations of a vector and the tensor suggests a mixing
of the scalar, vector and tensor sectors in inhomogeneous cosmology. Therefore, for a completely
rigorous treatment one should look at the quadratic action on the SdS background and take into
account the couplings of the modes.

 Furthermore, we make use of the slow-roll paradigm. This means that, even though strictly
speaking our results will be derived in SdS space, we shall assume that they hold in quasi de
Sitter space endowed with a small (decaying) black hole, provided one exacts the replacements:
$H\rightarrow H(t)$ and $M\rightarrow M(t)$. This is justified when both the Hubble parameter and
the black hole mass change adiabatically in time, in the sense that $\dot H\ll H_0^2$ ($\epsilon\ll
1$) and $\dot M\ll MH_0$.

\section{The propagators}
\label{PropagatorSection}

 To study scattering experiments, one typically calculates the S-matrix elements. In cosmology, on
the other hand, one is primarily interested in expectation values of operators with respect to some
definite vacuum state. For this purpose the {\tt in-in}, or
Schwinger-Keldysh~\cite{Schwinger:1960qe,Keldysh:1964ud},
formalism~\cite{Jordan:1986ug,Weinberg:2005vy,Prokopec:2010be} is suitable, in which time evolution
of an operator is described in terms of the perturbation theory based on the Keldysh propagator and
the {\tt in-in} vertices. Since here we are primarily interested in the spectrum of a scalar field
on SdS space, which can be obtained from any equal time two-point correlator, for our purpose it
suffices to calculate the corresponding Keldysh propagator.

 The Keldysh propagator is a $2\times 2$-matrix of the form,
\begin{align}
i\mathbf{G}(x;x') =
\left(\begin{array}{cc} iG_{++}(x;x') & iG_{+-}(x;x')
  \\
                 iG_{-+}(x;x') & iG_{--}(x;x') \end{array} \right)
\,,
\label{Keldysh propagator}
\end{align}
whose components are the Wightman functions $iG_{+-}$, $iG_{-+}$ and (anti-)time ordered Feynman
propagators $iG_{++}$,  $iG_{--}$, defined as,
\begin{align}
iG_F (x;x') \equiv\ &iG_{++}(x;x') =
        \langle\Omega|T\phi(x)\phi(x')|\Omega\rangle\qquad &(\mathrm{Feynman})
\\
&iG_{+-}(x;x')
= \langle\Omega|\phi(x')\phi(x)|\Omega\rangle\qquad &(\mathrm{Wightman})
\nonumber \\
&iG_{-+}(x;x')
 = \langle\Omega|\phi(x)\phi(x')|\Omega\rangle\qquad &(\mathrm{Wightman})
\nonumber \\
&iG_{--}(x;x')
= \langle\Omega|\overline{T}\phi(x)\phi(x')|\Omega\rangle
\qquad &(\mathrm{anti-Feynman})
\,,
\nonumber
\end{align}
where $|\Omega\rangle$ is a suitably chosen vacuum state. The time ordering is defined as
\begin{align}
T\phi(x)\phi(x')&= \theta(x_0-x_0')\phi(x)\phi(x')
                + \theta(x_0'-x_0)\phi(x')\phi(x)
\\
\overline{T}\phi(x)\phi(x') &= \theta(x_0-x_0')\phi(x')\phi(x)
                            + \theta(x_0'-x_0)\phi(x)\phi(x')
\,,
\nonumber
\end{align}
{\it i.e.}\ later times are to the left for $T$ and early times are to the left for $\overline{T}$.
The propagator $i\mathbf{G}$ satisfies the equation
\begin{equation} \label{GreensFunction}
\sqrt{-g(x)}\left(\square_x - m^2_\mathrm{eff}\right)i\mathbf{G}(x;x')
 = i\sigma^3\delta^4(x-x')
\,,
\end{equation}
where $\sigma^3$ is the Pauli matrix
\begin{equation}
\sigma^3 = \left(\begin{array}{cc} 1 & 0 \\ 0 & -1 \end{array} \right)
\end{equation}
and $m^2_\mathrm{eff}=d^2V(\phi_0)/d\phi_0^2$ is the effective mass-squared of the field, which in
the following we neglect. In slow-roll inflation $m_\mathrm{eff}$ can be expressed in terms of the
second slow-roll parameter $\eta_V=M_P^2V^{\prime\prime}/V$, {\it i.e.}\ $m^2_\mathrm{eff}=\eta_V
V(\phi_0)/M_P^2$, such that setting $m_{\rm eff}\rightarrow 0$ is equivalent to $\eta_V\rightarrow
0$.

\subsection{The de~Sitter case}
\label{dScase}

 In the de~Sitter case we can solve the equation of motion for the massless scalar
field~(\ref{eom:massless scalar in dS}--\ref{scalar dAlembertian}) explicitly. Taking advantage of
spatial homogeneity of de~Sitter space, the following mode decomposition of the free field is
convenient,
\begin{equation}
\phi(\eta,\vec{x})
= \int\frac{d^3\vec{k}}{(2\pi)^3}
  \left[u_k(\eta){\rm e}^{i\vec{k}\cdot\vec{x}}b(\vec{k}\,)
        + u^*_k(\eta){\rm e}^{-i\vec{k}\cdot\vec{x}}b^\dagger(\vec{k}\,)
  \right]
\,,
\label{field decomposition:modes}
\end{equation}
where $b(\vec{k}\,)$ and $b^\dagger(\vec{k}\,)$ are the annihilation and creation operators,
defined by $b(\vec{k}\,)|\Omega\rangle=0$ and by the commutation relation,
$[b(\vec{k}\,),b^\dagger(\vec{k}^\prime\,)] =(2\pi)^3\delta^3(\vec{k}-\vec{k}^\prime\,)$. The mode
functions $u_k(\eta)$ in~(\ref{field decomposition:modes}) satisfy the equation
\begin{equation}
\left(\partial_\eta^2+k^2-\frac{2}{\eta^2}\right)[au_k(\eta)]=0
\,.
\end{equation}
We obtained this result by making use of~(\ref{dAlembertian:dS}) and noting that $a(\eta) =
-1/(H_0\eta)$ implies $a^{\prime\prime}/a=2/\eta^2$. Imposing the boundary condition that the mode
functions behave like in the conformal vacuum in the asymptotic past yields
\begin{equation}
u_k(\eta)
 = \frac{1}{a}\frac{1}{\sqrt{2k}}\left(1-\frac{i}{k\eta}\right){\rm e}^{-ik\eta}
\,.
\label{mode function:dS}
\end{equation}
This equation, together with the condition $b(\vec{k}\,)|\Omega\rangle=0$ (for all $\vec{k}$),
defines the Bunch-Davies (BD) vacuum $|\Omega\rangle$. The fact that in the ultraviolet ($k/a\gg
H_0$) the BD vacuum minimizes the energy in the field fluctuations has led to the belief that this
vacuum represents a sensible physical choice~\cite{Birrell:1982ix} for the inflationary vacuum.
This is not so because in the infrared (where $k/a\ll H_0$) the BD vacuum yields infinite energy.
Namely, in the IR adiabaticity of the state is broken because the field couples strongly to the
expanding background, leading to abundant particle generation, having as a consequence strongly
enhanced infrared correlations. While this particle creation is very welcome in cosmology, since
the amplified vacuum fluctuations provide a beautiful explanation for the Universe's structure
formation, one has to take proper care to regulate the IR. One way of doing that is to replace the
BD vacuum by a more general state, characterized by the following generalization of the mode
functions~(\ref{mode function:dS}),
\begin{equation}
u_k(\eta)\rightarrow \tilde u_k(\eta)
  = \alpha(k)u_k(\eta) + \beta(k) u_k^*(\eta)
\,;\qquad |\alpha(k)|^2-|\beta(k)|^2 = 1
\,.
\label{mode function:dS:2}
\end{equation}
By suitably choosing $\beta(k)$, one can then make the infrared part of the vacuum state
finite~\cite{Vilenkin:1982wt}. A concrete working realization of this proposal has been
investigated in Refs.~\cite{Janssen:2009nz,Koivisto:2010pj}. Alternatively, one can remove the
infrared problems by placing the Universe in a large comoving box of size $L$. This leads to a
discretized reciprocal (momentum) space $\vec k=\vec n k_0$ [$\vec n=(n_1,n_2,n_3)$, with $n_i$
integers], with the comoving lattice size $k_0=2\pi/L$. Since $k_0$ corresponds to the minimum
allowed momentum, this cures the infrared problem simply by disallowing the deeply infrared modes.
In the limit when $L\rightarrow\infty$, the lattice constant $k_0\rightarrow 0$, and the sum over
the momenta can be replaced with increasing accuracy by an integral, which has $k_0$ as the IR
cut-off, thus regulating the infrared. From the physical point of view, it is natural to associate
this cut-off with the scale of the Hubble horizon at the beginning of inflation, thereby
eliminating modes that stretch beyond the Hubble radius. One way of implementing this, is to take
the spatial topology of the universe to be compact, {\it e.g.}\ a torus, as discussed
in~\cite{Tsamis:1994ca}.

 To see how the regularization procedure works in practice, we shall now calculate the regulated
Feynman propagator in de Sitter space. In order to do that, we need to relate the direct space
propagator to its mode functions~(\ref{mode function:dS}). Because de Sitter space is spatially
homogeneous, it is convenient to write the components $i\Delta_{ab}(x;x')$ of the Keldysh
propagator~(\ref{Keldysh propagator}) for de Sitter space in terms of its Fourier space
counterparts,
\begin{align}
i\Delta_{ab}(x;x') = \int \frac{d^3\vec{k}}{(2\pi)^3}
  {\rm e}^{i\vec{k}\cdot(\vec{x}-\vec{x}')}
       i\Delta_{ab}(k,\eta,\eta')
\,.
\label{propagator:Fourier transform}
\end{align}
Making use of Eqs.~(\ref{field decomposition:modes}) and~(\ref{mode function:dS}), one finds for
the momentum space propagators,
\begin{align}
&i\Delta_{+-}(k,\eta,\eta^\prime) = u_k^*(\eta)u_k(\eta^\prime)
\nonumber\\
&i\Delta_{-+}(k,\eta,\eta^\prime) = u_k(\eta)u_k^*(\eta^\prime)
 = (i\Delta_{+-}(k,\eta,\eta^\prime))^*
\nonumber \\
&i\Delta_{++}(k,\eta,\eta^\prime)
   = \Theta(\eta-\eta^\prime)i\Delta_{-+}(k,\eta,\eta^\prime)
        + \Theta(\eta^\prime-\eta)i\Delta_{+-}(k,\eta,\eta^\prime)
\label{propagators:dS:momentum}
\\
&i\Delta_{--}(k,\eta,\eta^\prime)
    = \Theta(\eta-\eta^\prime)i\Delta_{+-}(k,\eta,\eta^\prime)
      + \Theta(\eta^\prime-\eta)i\Delta_{-+}(k,\eta,\eta^\prime)
\,.
\nonumber
\end{align}
 The corresponding spectrum ${\cal P}^{dS}_\phi$, defined by
\begin{equation}
 \langle\Omega|\phi(\vec x,\eta)\phi(\vec x^\prime,\eta)|\Omega\rangle =
   \int \frac{dk}{k}  {\cal P}^{dS}_\phi(k,\eta)
      \frac{\sin\left(k\|\vec x-\vec x^{\,\prime}\|\right)}{k\|\vec x-\vec x^{\,\prime}\|}
\,,
\label{power spectrum:dS}
\end{equation}
is obtained straightforwardly from the equal time limit ($\eta^\prime\rightarrow\eta$) of the
propagator,
\begin{equation}
{\cal P}^{dS}_\phi(k,\eta)
  = \frac{k^3}{2\pi^2}i\Delta_{+-}(k,\eta,\eta)
  = \frac{H_0^2}{4\pi^2}\left(1+k^2\eta^2\right)
\,.
\label{dSSpectrum}
\end{equation}
It is scale invariant at future infinity, $\eta\rightarrow 0$.

 Based on~(\ref{propagators:dS:momentum}) one can calculate the position space de Sitter propagator by
performing the momentum integral~(\ref{propagator:Fourier transform}) over $k\geq k_0$. The
resulting Feynman propagator is~\cite{Tsamis:1993ub,Prokopec:2003tm},
\begin{equation}
 i\Delta_F(x;x^\prime)
= \frac{H_0^2}{4\pi^2}\bigg(\frac{\eta\eta^\prime}{\Delta x^2}
  - \frac{1}{2}\Big[\log\left(k_0^2\Delta x^2\right)+2(\gamma_\mathrm{E}-1)\Big] + \mathcal{O}(k_0)\bigg)
\,,
\label{de Sitter propagtor:direct space}
\end{equation}
where $\Delta x^2 = -(|\eta-\eta^\prime|-i\epsilon)^2+\|\vec x-\vec x^\prime\,\|^2$ is the
conformal space distance function. Two comments are in order. Firstly, apart from the standard
Hadamard contribution $\propto 1/\Delta x^2$, which is singular on the lightcone (on-shell) and
quickly decays off-shell, due to rapid particle production in de Sitter space, the de Sitter
propagator~(\ref{de Sitter propagtor:direct space}) acquires a logarithmic term which contributes
both within the past and future light cones. Secondly, the logarithm grows without a limit as
$k_0\rightarrow 0$. This is a manifestation of the IR singularity of the Bunch-Davies vacuum. We
will see below that a black hole in de Sitter space `sees' this logarithmic singularity in the
corrected SdS propagator as a logarithmic singularity in the mixed space propagator and hence also
in the (mixed space) SdS spectrum. This dependence on the IR regulator poses a unique opportunity
to investigate the black hole contribution to the spectrum dependent on the IR regularization. In
this paper we choose the comoving box regulator primarily because of its simple implementation, but
one would certainly benefit from studying other IR regularization schemes.

\subsection{The Schwarzschild-de~Sitter case}

 For the case when a primordial black hole is present in de Sitter space we shall derive only the
first order correction in $\mu$ to the Schwinger-Keldysh propagator. For this we write
\begin{equation} \label{FullSdSpropagator}
 \square = \square^{dS}+\delta\square
\,; \qquad
 \sqrt{-g} = \sqrt{-g_{dS}} + \delta\sqrt{-g}
\,;\qquad
 i\mathbf{G} = i\mathbf{\Delta} + i\delta\mathbf{G}
\,,
\end{equation}
with $i\mathbf{\Delta}$ being the propagator on de~Sitter space~(\ref{propagator:Fourier
transform}--\ref{propagators:dS:momentum}), (\ref{de Sitter propagtor:direct space}). By plugging
this into~(\ref{GreensFunction}) we find that the correction to $i\delta\mathbf{G}$ satisfies:
\begin{align} \label{CorrectedGreensFunction}
&\sqrt{-g_{dS}(x)}\left(\square^{dS}_x - m^2_\mathrm{eff}\right)
         i\delta\mathbf{G}(x;x') = \\
&\hskip 3cm -\left(\sqrt{-g_{dS}(x)}\delta\square_x i\mathbf{\Delta}(x;x')
  + i\frac{\delta\sqrt{-g(x)}}{\sqrt{-g_{dS}(x)}}\sigma^3\delta^4(x-x')\right)
\,. \nonumber
\end{align}
Note that $\delta\sqrt{-g}$ is only $\mathcal{O}(\mu^6)$, and we will neglect it from now on. It
follows that
 \begin{align}
i\delta\mathbf{G}(x;x')
  &= i \int d^4 x'' \sqrt{-g_{dS}(x'')} i\mathbf{\Delta}(x;x'')\sigma^3
           \delta\square_{x''} i\mathbf{\Delta}(x'';x')
\,.
\label{CorrectedGreensFunction:solution}
\end{align}
This solution of~(\ref{CorrectedGreensFunction}) is given only up to a homogeneous solution of the
d'Alembertian operator in~(\ref{CorrectedGreensFunction}). The unique propagator
in~(\ref{CorrectedGreensFunction:solution}) is obtained upon specifying the boundary conditions for
the mode functions, or equivalently, for the vacuum state. Here the unperturbed vacuum state is
chosen to be the (pure) Bunch-Davies vacuum of de Sitter space, whereby the deep infrared modes are
removed by placing the Universe in a comoving box, as explained in section~\ref{dScase}. But we are
still free to add a homogeneous solution to the Feynman propagator (resulting in a mixed state),
which has to be added to take Hawking radiation into account. The light black holes that we
consider do indeed emit Hawking radiation but a simple estimate~\footnote{From Table 1
in~\cite{Candelas:1980zt} one can in principle obtain an explicit expression for the asymptotic
form of the (renormalized) propagator for Hawking radiation in the Unruh vacuum, which is the
physically relevant state in this case. However, a rigorous treatment of the effect of Hawking
radiation requires the knowledge of the asymptotic solutions to the confluent Heun's equation which
determine the radial mode functions but these asymptotic solutions are, to our knowledge, not
known. A detailed analysis which could support our conjecture in Eq.~\eqref{EmissionRates} is
beyond the scope of this work.} suggests that this does not change the spectrum of scalar
fluctuations at the leading order in the perturbation parameter $\mu$. Namely, comparing the
emission rate per Hubble time $t_H$ and Hubble surface area $A_H$ for inflaton fluctuations and
Hawking radiation, we find that
\begin{equation}\label{EmissionRates}
\frac{E}{A_H t_H}\Bigg|_\phi \sim H_0^4\,,\qquad \frac{E}{A_H t_H}\Bigg|_{Haw.\ rad.} \sim
\dot{M}H_0^2 \left(\frac{R_S}{R_H}\right)^4 \sim \mu^6 H_0^4\,,
\end{equation}
where the factor $(R_S/R_H)^4$ accounts for the redshift of Hawking radiation from the time of its
creation, when the typical wavelength is of the order of the Schwarzschild radius, to the time when
the amplitude freezes out, when the wavelength is of the order of the Hubble radius. This estimate
holds in the case when the scalar field is light (or massless), $m \ll H_0$. Note that there is no
enhancement in the second equation of~\eqref{EmissionRates} by the total number of degrees of
freedom $g_*$ ({\it cf.} Eq.~\eqref{EvaporationTime}) because only non-conformally coupled matter
fields (scalars and tensors) are relevant for the radiation from the black hole as their amplitude
freezes out after exiting the Hubble radius. From~\eqref{EmissionRates} we find that the ratio of
the emission rates is suppressed as $\sim \mu^6$ since $\dot{M}\sim M_P^4/M^2 \varpropto \mu^{-6}$
and $(R_S/R_H)^4 \sim \mu^{12}$. The emission of very massive particles, $m^{-1}< GM$, is
exponentially suppressed, making them irrelevant for the above estimate. In the intermediate
regime, $GM < m^{-1}< H_0^{-1}$, the ratio of the emission rates is $\sim \mu^3 (H_0/m)$. This
dimensional analysis indicates that Hawking radiation does not contribute at the leading order
which is $\mu$ or $\mu^3$, depending on the scale, as can be seen in~\eqref{ImJZero}
and~\eqref{spectrum:BH correction} arising from the homogeneous contribution to the propagator.

Since $i\delta\mathbf{G}(x;x')\rightarrow 0$ for $M\rightarrow 0$, our modified vacuum state
reduces to the Bunch-Davies vacuum of de~Sitter space in this limit, as it should. When
Eq.~(\ref{CorrectedGreensFunction:solution}) is written in its component form we get,
\begin{align}
i\delta G_{ab}(x;x')
&= i \sum_{c=+,-}c \int d^4 x'' \sqrt{-g_{dS}(x'')}
 i\Delta_{ac}(x;x'')\delta\square_{x''} i\Delta_{cb}(x'';x')
\,,
\end{align}
with $a,b=+,-$. Writing the propagator $i\Delta$ in momentum space~(\ref{propagator:Fourier
transform}--\ref{propagators:dS:momentum}) this becomes
\begin{align}
&i\delta G_{ab}(x;x') \label{CorrectedGreensFunctionPositionSpace} \\
&\quad = -\frac{4i\mu^3}{3H_0^2}\int \frac{d^3 \vec{k}d^3 \vec{k}'}{(2\pi)^6}\int_{\eta_0}^0 d\eta''\eta''\left(\sum_{c=+,-}c\ i\Delta_{ac}(k,\eta,\eta'')i\Delta_{cb}(k',\eta'',\eta')\right) {\rm e}^{i(\vec{k}\cdot\vec{x}-\vec{k}'\cdot\vec{x}')} \nonumber \\
&\qquad \times \int_{-\mu \eta''}^\infty \frac{dr''}{r''} \int_{-1}^1 d\cos\theta''\int_0^{2\pi} d\phi'' \left(-k'^2+3\frac{(\vec{k}'\cdot\vec{x}'')^2}{r''^2}\right){\rm e}^{i(\vec{k}'-\vec{k})\cdot\vec{x}''}
\,. \nonumber
\end{align}
It turns out that it is easiest to evaluate a (double) momentum space version of $i\delta G_{ab}$.
To do that, we first introduce the momenta associated with the positions $x$ and $x'$,
\begin{equation}
i\delta G_{ab}(\vec{p_1},\vec{p_2},\eta,\eta') = \int d^3\vec{x}\, d^3\vec{x}'\, i\delta
G_{ab}(\vec{x},\vec{x}',\eta,\eta') {\rm e}^{-i\vec{p_1}\cdot \vec{x}}{\rm e}^{-i\vec{p_2}\cdot
\vec{x}'},
\end{equation}
and, next, relative and average coordinates in position and momentum space,
$\vec{r}=\vec{x}-\vec{x}'$, $\vec{y}=(\vec{x}+\vec{x}')/2$ and $\vec{p}=\vec{p}_1+\vec{p}_2$,
$\vec{q}=(\vec{p}_1-\vec{p}_2)/2$. This yields
\begin{eqnarray} \label{CorrectedGreensFunctionMomentumSpace}
i\delta G_{ab}(\vec{p},\vec{q},\eta,\eta')
 &=& \int d^3\vec{y}\, d^3\vec{r}\, i\delta G_{ab}(\vec{x},\vec{x}',\eta,\eta')
  {\rm e}^{-i\vec{p}\cdot \vec{y}}{\rm e}^{-i\vec{q}\cdot \vec{r}} \\
&&\hskip -2.4cm
 = -\frac{16\pi i\mu^3}{3H_0^2}k'^2 (3\cos^2\widetilde{\theta}-1)
    \sum_{c=+,-} c\left(J_{ac,cb}(0; \vec{p},\vec{q},\eta,\eta')
                -J_{ac,cb}(\eta_0; \vec{p},\vec{q},\eta,\eta')\right)
\,, \nonumber
\end{eqnarray}
where here and in what follows $\vec{k}=\vec{q}+\frac{1}{2}\vec{p}$,
$\vec{k'}=\vec{q}-\frac{1}{2}\vec{p}$ and $\widetilde{\theta} = \sphericalangle
(\vec{k}'-\vec{k},\vec{k}') = \sphericalangle (-\vec{p},\vec{q}-\frac{1}{2}\vec{p})$. Moreover, we
defined
\begin{align}
 J_{ab,cd}(\eta''; \vec{p},\vec{q},\eta,\eta')
 \!=\! \int\! d\eta''\eta''i\Delta_{ab}(k,\eta,\eta'')
   i\Delta_{cd}(k',\eta'',\eta')
     \left(\frac{\cos(\mu p\eta'')}{\mu^2p^2\eta''^2}
           -\frac{\sin(\mu p\eta'')}{\mu^3p^3\eta''^3}
      \right)
.
\end{align}
The details of the derivation can be found in Appendix~B. From
Eq.~(\ref{CorrectedGreensFunctionPositionSpace}) and relations~(\ref{Theta functions:relations})
for the step functions, we find that the corrections to the Feynman and anti-Feynman propagators
obey the standard time-ordering relations
\begin{align}
&i\delta G_{++}(\vec{p},\vec{q},\eta,\eta') = \Theta(\eta-\eta')i\delta G_{-+}(\vec{p},\vec{q},\eta,\eta') + \Theta(\eta'-\eta)i\delta G_{+-}(\vec{p},\vec{q},\eta,\eta') \label{FeynmanPropagators}\\
&i\delta G_{--}(\vec{p},\vec{q},\eta,\eta') = \Theta(\eta-\eta')i\delta G_{+-}(\vec{p},\vec{q},\eta,\eta') + \Theta(\eta'-\eta)i\delta G_{-+}(\vec{p},\vec{q},\eta,\eta') \label{FeynmanPropagatorsAnti} \,.
\end{align}
In addition, we have
\begin{equation} \label{Wightman2}
  i\delta G_{-+}(\vec{p},\vec{q},\eta,\eta')
  = \Big(i\delta G_{+-}(\vec{p},\vec{q},\eta,\eta')
    \Big)^*
\,.
\end{equation}
Therefore, in order to fully reconstruct the black-hole-corrected Keldysh propagator,we only have
to determine $i\delta G_{+-}$, for which we need to know only $J_{+-,+-}$ and $J_{+-,-+}$:
\begin{align}
&J_{+-,+-}(\eta''; \vec{p},\vec{q},\eta,\eta') = \frac{H_0^4\eta\eta' {\rm e}^{i(k\eta-k'\eta')}}{4kk'(\mu p)^4}\left(1+\frac{i}{k\eta}\right)\left(1-\frac{i}{k'\eta'}\right) {\rm e}^{-i(k-k')\eta''} \label{FinalJs} \\
&\quad \times\Bigg\{\bigg[-\frac{(\mu p)^2}{kk'}+\frac{i(\mu p)^2(k-k')\eta''-\frac{(\mu p)^4}{kk'}}{(k-k')^2-(\mu p)^2}+\frac{2(\mu p)^4}{((k-k')^2 -(\mu p)^2)^2}\bigg]\cos(\mu p\eta'') \nonumber \\
&\; + \bigg[\frac{\mu p}{kk'\eta''}-\frac{(\mu p)^3\eta''+\frac{i(\mu p)^3(k-k')}{kk'}+i(\mu p)(k\!-\!k')}{(k-k')^2-(\mu p)^2}+\frac{2i(\mu p)^3(k-k')}{((k\!-\!k')^2-(\mu p)^2)^2}\bigg]\sin(\mu p\eta'')\Bigg\} \nonumber \\ \nonumber \\
&J_{+-,-+}(\eta''; \vec{p},\vec{q},\eta,\eta') = \frac{H_0^4\eta\eta' {\rm e}^{i(k\eta+k'\eta')}}{4kk'(\mu p)^4}\left(1+\frac{i}{k\eta}\right)\left(1+\frac{i}{k'\eta'}\right) {\rm e}^{-i(k+k')\eta''} \label{FinalJsB}\\
&\; \times\Bigg\{\bigg[\frac{(\mu p)^2}{kk'}+\frac{i(\mu p)^2(k\!+\!k')\eta''+\frac{(\mu p)^4}{kk'}}{(k+k')^2-(\mu p)^2}+\frac{2(\mu p)^4}{((k\!+\!k')^2 -(\mu p)^2)^2}\bigg]\cos(\mu p\eta'') \nonumber \\
&\; - \bigg[\frac{\mu p}{kk'\eta''}+\frac{(\mu p)^3\eta''-\frac{i(\mu p)^3(k\!+\!k')}{kk'}+i(\mu p)(k+k')}{(k+k')^2-(\mu p)^2}-\frac{2i(\mu p)^3(k\!+\!k')}{((k\!+\!k')^2-(\mu p)^2)^2}\bigg]\sin(\mu p\eta'')\Bigg\}. \nonumber
\end{align}
\begin{comment}
$J_{+-,-+}$ is obtained from $J_{+-,+-}$ by the formal substitution $k'\rightarrow -k'$,
\begin{equation}
J_{+-,-+}(\eta''; \vec{p},\vec{q},\eta,\eta') = -J_{+-,+-}(\eta''; \vec{p},\vec{q},\eta,\eta')\Big|_{k'\rightarrow -k'}\,.
\end{equation}
\end{comment}
%
From Eq.~(\ref{CorrectedGreensFunctionMomentumSpace}) we get,
\begin{align}
i\delta G_{+-}(\vec{p},\vec{q},\eta,\eta')
 &= -\frac{16\pi i\mu^3}{3H_0^2}k'^2(3\cos^2\widetilde{\theta}-1)
  \big[J^*_{+-,-+}(\eta)-J_{+-,+-}(\eta)
\label{Wightman1}
\\
&\quad + J_{+-,+-}(\eta') - J_{+-,-+}(\eta')
        + J_{+-,-+}(\eta_0) - J^*_{+-,-+}(\eta_0) \big]
\,, \nonumber
\end{align}
where here, for brevity, we wrote $J_{ab,cd}(\eta'')\equiv J_{ab,cd}(\eta'';
\vec{p},\vec{q},\eta,\eta')$. In order to get the complete SdS propagator, we still need to add the
de Sitter propagator to the correction~(\ref{Wightman1}), which in the double Fourier space reads,
\begin{equation}
i\Delta_{+-}(\vec{p},\vec{q},\eta,\eta')
% &=& u_k^*(\eta)u_k(\eta^\prime)\int d^3y{\rm e}^{-i\vec p\cdot\vec k}
%\nonumber\\
 = \frac{H_0^2\eta\eta^\prime}{2q}\Big(1+\frac{i}{q\eta}\Big)
     \Big(1-\frac{i}{q\eta^\prime}\Big){\rm e}^{-iq(\eta-\eta^\prime)}
         (2\pi)^3\delta^3(\vec p\,)
\,.
\label{Wightman:dS:double}
\end{equation}
Together with Eqs.~(\ref{FeynmanPropagators}--\ref{FinalJs}) and~(\ref{Wightman:dS:double}),
relation~(\ref{Wightman1}) completely determines the desired SdS propagator in the limit of a small
black hole mass, and it constitutes our main result. An interesting feature of the
propagator~(\ref{Wightman1}) is that, due to causality, it does not contain any information from
future infinity ($\eta''=0$). Notice that, although $\mu \ll 1$, we have not expanded the factors
$\sin(\mu p\eta'')$ and $\cos(\mu p\eta'')$ in powers of $\mu$ in
Eqs.~(\ref{FinalJs}--\ref{FinalJsB}) since one might still want to consider momenta $\vec{p}$ with
$-\mu p \eta'' \gtrsim 1$. Consequently, even though our original expansion parameter was $\mu^3$,
the propagator correction is formally suppressed only as $\mu^1\propto M^{1/3}$, thus, as a
fractional power of the black hole mass. Finally, it is worth noting that the pole of $J_{+-,-+}$
at $k+k'=\mu p$ in~(\ref{FinalJsB}) is not physically realized for any $\mu<1$, because $k+k'=
\|\vec{q}+\vec{p}/2\| + \|\vec{q}-\vec{p}/2\|
> p$.

\section{The power spectrum}
\label{PowerSpectrumSection}

\subsection{Double momentum space representation}

 In section~\ref{dScase} we derived the spectrum~(\ref{dSSpectrum}) for a massless scalar field on
de~Sitter space. The inhomogeneous case with a primordial black hole is far less trivial to deal
with, mainly because out of the 10 symmetries (Killing vectors) of de~Sitter space only three
symmetries remain in Schwarzschild-de~Sitter space. (Recall that the homogeneous cosmology of
slow-roll inflation, radiation and matter era has six symmetries.)

 In principle, one has to rederive the gauge invariant combinations of the fields, such as the
Sasaki-Mukhanov field~(\ref{SakakiMukhanov}) $\mathcal{R}$, for the inhomogeneous background and
determine their power spectra. But we have seen in Eq.~(\ref{SasakiMukhanovFieldForSdS}) that
$\mathcal{R}$ is approximately gauge invariant for weak breaking of translational symmetry.
Similarly, the graviton contribution can be neglected, since its power spectrum is expected to be
equally suppressed, Eq.~(\ref{rParameterForGravitonSpectrum}).

\begin{figure}
\centering
\includegraphics[scale=0.7]{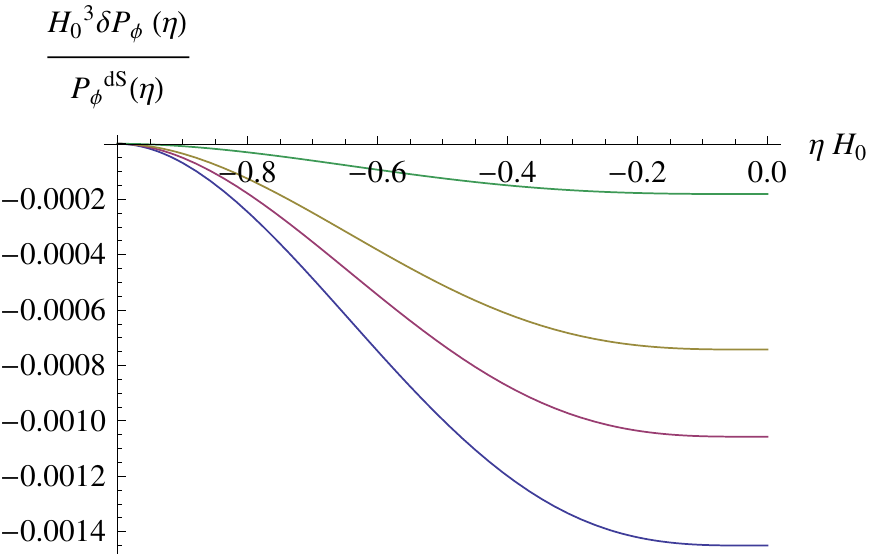}
\includegraphics[scale=0.7]{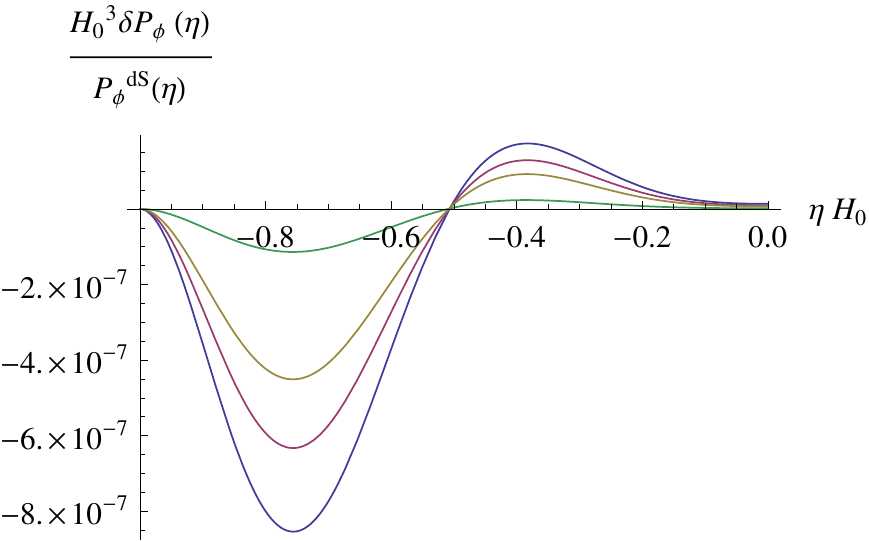}
\caption{\footnotesize The dependence on $\eta$ of the correction to the power spectrum in (double)
momentum space, $\delta \mathcal{P}_\phi(\vec{p},\vec{q},\eta)=(q^3/(2\pi^2)) i\delta
G_{+-}(\vec{p},\vec{q},\eta,\eta)$, rescaled by the de~Sitter spectrum $\mathcal{P}_\phi
(\vec{q},\eta) = (H_0^2/(4\pi^2)) (1+q^2\eta^2)$, is presented here for fixed momenta $\vec p$,
$\vec q$ and different values of $\mu$ (from bottom to top: $\mu=0.1$ (blue curve), $\mu=0.09$
(pink), $\mu=0.08$ (yellow) and $\mu=0.05$ (green)). The correction is zero at the initial
hypersurface, $\eta=-H_0^{-1}$. For small $\eta$ the spectrum is well described by an expansion to
order $\eta^3$. The linear order vanishes. It approaches a non-zero value in the limit
$\eta\rightarrow 0$. Left panel: $p/H_0=q/H_0=1$ and $\sphericalangle(\vec p,\vec q) = 0$, right
panel: $p/H_0=10$, $q/H_0=1/10$ and $\sphericalangle(\vec p,\vec q) = \pi/2$. Note that in this and
in the subsequent plots the relative correction to $\mathcal{P}_\phi$ is shown but this coincides
with the relative correction to $\mathcal{P_R}$ because of Eq.~\eqref{relativeCorrSpectra}.}
\label{fig: EtaDependenceOfSpectrum}
\end{figure}

\begin{figure}
\centering
\includegraphics[scale=0.7]{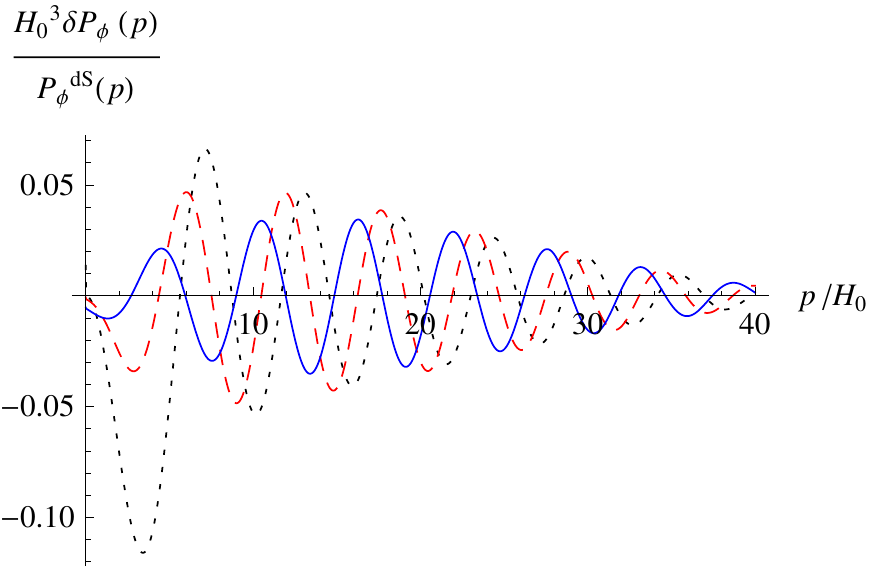}
\includegraphics[scale=0.7]{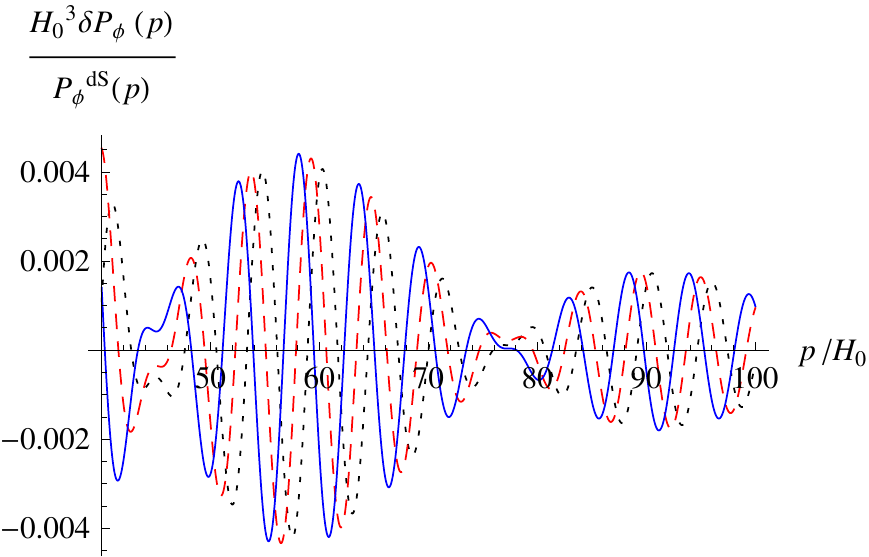}
\caption{\footnotesize In this plot we show the spectrum in (double) momentum space close to the
singular region for $\mu=1/10$ and $\eta=0$. For this we choose the angle between $\vec p$ and
$\vec q$ to be $\theta=0.9\pi$ and $q=p/2$ (black curve, dotted), $q=p/2+H_0$ (red, dashed) and
$q=p/2+2H_0$ (blue, solid). We observe oscillatory behavior with decaying amplitude. The spectrum
diverges for $\vec{q}=-\vec{p}/2$.} \label{fig: SpectrumInDoubleMomentumSpace}
\end{figure}

 With these assumptions the correction to the spectrum from the black hole can be determined from
$i\delta G_{+-}$ by taking the equal time limit $\eta'\rightarrow \eta$ of the
propagator~(\ref{Wightman1}),
\begin{eqnarray}
i\delta G_{+-}(\vec{p},\vec{q},\eta,\eta)
 &=& \frac{32\pi\mu^3}{3H_0^2}k'^2(3\cos^2\widetilde{\theta}-1)
\label{spectrum:BH correction}
\\
&&\times\,\big[\mathrm{Im}\, J_{+-,-+}(\eta_0;\vec{p},\vec{q},\eta,\eta)
      - \mathrm{Im}\, J_{+-,-+}(\eta;\vec{p},\vec{q},\eta,\eta)
  \big]
\,.
\nonumber
\end{eqnarray}
The $\eta$ dependence of the corrected power spectrum $\delta \mathcal{P}_\phi (\vec p, \vec q,
\eta)$ is displayed in Fig.~\ref{fig: EtaDependenceOfSpectrum}. Note that, because of
Eq.~\eqref{spectrum:comoving gauge}, the relative correction to the spectrum of inflaton
fluctuations coincides with the correction to the spectrum of $\mathcal{R}$,
\begin{equation} \label{relativeCorrSpectra}
\frac{\delta \mathcal{P_R}}{\mathcal{P_R}^\mathrm{qdS}} = \frac{\delta
\mathcal{P}_\phi}{\mathcal{P}_\phi^\mathrm{dS}}\,,
\end{equation}
where, using the slow-roll approximation $\epsilon \ll 1$, the spectrum $\mathcal{P_R}$ is the one
for quasi de Sitter space (qdS). This should be taken into account when physically interpreting the
plots in the subsequent Figs.~\ref{fig: EtaDependenceOfSpectrum}--\ref{fig: SpectrumIsotropic}
because it is the fluctuations in $\mathcal{R}$ that are directly related to the observable
temperature fluctuations (this will be further explained in the discussion section, {\it cf.}\
also~\eqref{temperature corr fn}). The correction to the spectrum $\delta \mathcal{P_R}$ vanishes
at the initial hypersurface $\eta_0=-H_0^{-1}$ and approaches a non-zero value at $\eta=0$. This
means that by the end of inflation an imprint of a small black hole on the spectrum will remain.

 For a homogeneous background the propagator in (double) momentum space contains a delta-peak in the
momentum $\vec{p}$ that is associated with the average of the positions. This is seen explicitly in
the de~Sitter case from Eq.~(\ref{Wightman:dS:double}). In the case of a small inhomogeneity we
find a power law divergence at $\vec k = \vec q+\vec p/2 = 0$. The behavior of the spectrum close
to this singularity is shown in Fig.~\ref{fig: SpectrumInDoubleMomentumSpace}.

 This observation suggests that an expansion in powers of $\vec p\cdot \vec q$ and $p$, which is the
basis of the analysis in Ref.~\cite{Carroll:2008br} is inappropriate for the complete analysis of
small black holes in inflation, and, in what follows, we shall not make use of this expansion.
Instead, we shall analyze the mixed space spectrum without making an expansion in powers of $\vec
p$.

 From Eqs.~(\ref{Wightman1}) and~(\ref{FinalJs}) it follows that $i\delta G_{+-} = 0$ at the initial
hypersurface ($\eta=\eta_0$). In other words, we consider a primordial black hole that was created
at a time $\eta=\eta_0$, and we study how it perturbs scalar quantum fluctuations during the
subsequent inflationary period.

 For $\mu\ll 1$ also $\mu p \ll k+k'$ holds and we can expand $J_{+-,-+}$ in~(\ref{FinalJs}) to get
\begin{align}
&J_{+-,-+}(\eta_0;\vec{p},\vec{q},\eta,\eta) = \frac{H_0^4 {\rm e}^{i(k+k')(\eta-\eta_0)}}{4kk'(\mu p)^2}\left(\eta^2-\frac{1}{kk'}+\frac{i(k+k')\eta}{kk'}\right) \\
&\qquad \times\bigg\{\left(\frac{1}{kk'}+\frac{i\eta_0}{k+k'}\right)\left(\cos(\mu p\eta_0)-\frac{\sin(\mu p\eta_0)}{\mu p\eta_0}\right) \nonumber \\
&\qquad\quad + (\mu p)^2\bigg[\left(\frac{1}{kk'(k+k')^2}+\frac{2}{(k+k')^4}+\frac{i\eta_0}{(k+k')^3}\right)\cos(\mu p\eta_0) \nonumber \\
&\qquad\quad + \left(\frac{i\eta_0}{kk'(k+k')}-\frac{\eta_0^2}{(k+k')^2}+\frac{i\eta_0}{(k+k')^3}\right)\frac{\sin(\mu p\eta_0)}{\mu p\eta_0}\bigg] \bigg\}, \nonumber
\end{align}
where we kept $\sin$ and $\cos$ unexpanded. From this expression we see that there is a
contribution that remains finite at future infinity, {\it i.e.}\ at the end of inflation, meaning
that this correction to the spectrum is propagated through the radiation and matter dominated
epochs of the Universe. Taking the limit $\eta\rightarrow 0$ one finds
\begin{align} \label{ImJZero}
&\mathrm{Im}\, J_{+-,-+}(\eta_0;\vec{p},\vec{q},\eta,\eta) =
\\
& -\frac{H_0^4}{4k^2k'^2(\mu p)^2} \Bigg\{\left[\frac{\eta_0\cos((k+k')\eta_0)}{k+k'}-\frac{\sin((k+k')\eta_0)}{kk'}\right]\left(\cos(\mu p\eta_0)-\frac{\sin(\mu p\eta_0)}{\mu p\eta_0}\right) \nonumber
\\
& + (\mu p)^2\bigg[\frac{\eta_0\cos((k\!+\!k')\eta_0)}{(k+k')^3}
                 \cos(\mu p\eta_0) + \left(\frac{1}{kk'}
               + \frac{1}{(k\!+\!k')^2}\right)
                \frac{\eta_0\cos((k\!+\!k')\eta_0)}{(k+k')}
                    \frac{\sin(\mu p\eta_0)}{\mu p\eta_0}
\nonumber \\
& - \left(\frac{1}{kk'} + \frac{2}{(k+k')^2}\right)
           \frac{\sin((k+k')\eta_0)}{(k+k')^2} \cos(\mu p\eta_0)
   + \frac{\eta_0^2\sin((k+k')\eta_0)}{(k+k')^2}
                   \frac{\sin(\mu p\eta_0)}{\mu p\eta_0} \bigg]
 \Bigg\}
\nonumber
\\
& + \mathcal{O}(\eta)
\nonumber
\,.
\end{align}
Note that $\mathrm{Im}\, J_{+-,-+}(\eta;\vec{p},\vec{q},\eta,\eta) = \mathcal{O}(\eta)$. Therefore,
this contribution to the propagator is subdominant at late times and becomes completely negligible
by the end of inflation. Since we are primarily interested in late time cosmology, from now on we
shall not consider these terms. However, we should keep in mind that these terms become
increasingly important at early times when $\eta$ approaches $\eta_0$, since they guarantee that
$i\delta G_{ab}\rightarrow 0$ when $\eta\rightarrow \eta_0$.

\subsection{Mixed space representation}\label{MixSpaceRepSection}

 Rather than working with the momentum $\vec{p}$, we shall mainly consider the mixed space
propagator by means of a Fourier transformation of~(\ref{Wightman1}) with respect to $\vec{p}$. The
mixed space propagator $i\delta G_{+-}(\vec{y},\vec{q},\eta,\eta^\prime)$ is a function of the
relative momentum $\vec q$, the average position $\vec{y}=(\vec{x}+\vec{x}')/2$ and the times
$\eta$ and $\eta^\prime$. This makes the physical interpretation easier, because the corresponding
equal time statistical propagator
\begin{equation}
\delta F(\vec{y},\vec{q},\eta,\eta)
    = \mathrm{Re}\, i\delta G_{+-}(\vec{y},\vec{q},\eta,\eta)
\end{equation}
is closely related to the Boltzmann distribution function $\delta f$ ($\delta f \simeq q \delta
F$), \textit{cf}.\ \cite{Garbrecht:2002pd}, and hence allows for a simple statistical
interpretation as the phase space density of particles with momentum $\vec q$ at position $\vec y$
at time $\eta$. Likewise, the corresponding spectrum $\delta{\cal P}=[q^3/(2\pi^2)]\delta F$ can be
given an analogous simple statistical interpretation. The dependence on $\vec{p}$ as well as on
$\vec{y}$ in the mixed space propagator signals a breakdown of translational invariance.
Furthermore, the mixed space representation is advantageous also because the power law divergence
at $\vec p = - 2\vec q$, that plagues the double momentum space propagator, will become a mild
logarithmic divergence in the mixed space propagator, whose origin is the IR divergence of the BD
vacuum. From the above discussion we know that this divergence is regulated when the IR of the de
Sitter state is made finite. Just as we have regulated the de Sitter propagator~(\ref{de Sitter
propagtor:direct space}), we shall regulate this divergence by placing the Universe in a comoving
box of size $L=2\pi/k_0$. The mixed space propagator can then be written as
\begin{align}
&i\delta G_{+-}(\vec{y},\vec{q},\eta,\eta;k_0) = \int_{k\geq k_0} \frac{d^3\vec{p}}{(2\pi)^3}i\delta G_{+-}(\vec{p},\vec{q},\eta,\eta){\rm e}^{i\vec{p}\cdot \vec{y}} \\
&\qquad\qquad\qquad\qquad = \frac{32\pi\mu^3}{3H_0^2}
   \int_{k\geq k_0}\frac{d^3\vec{p}}{(2\pi)^3}
       k'^2(3\cos^2\widetilde{\theta}-1)
                 \mathrm{Im}\, J_{+-,-+}(\eta_0;\vec{p},\vec{q},\eta,\eta)
                   {\rm e}^{i\vec{p}\cdot \vec{y}}
\,.
\nonumber
\end{align}
An inspection of this integral shows that it is indeed logarithmically divergent in the IR when
$\vec p = - 2\vec q$, where $k=0$. The divergence is regulated by imposing $k\geq k_0$. The terms
in the integrand that cause this logarithmic divergence go as $1/k^3$ for small $k$ in $\mathrm{Im}
J_{+-,-+}$ (the integral is convergent at $\vec p = 2\vec q$ where $k^\prime=0$). Let us split
$i\delta G_{+-}$ into an IR finite part and an IR divergent part. For this we choose the $z$-axis
to point in the direction of $\vec{q}$ and keep $\vec{p}$ general, {\it i.e.}\ $\vec{q}=q(0,0,1)$
and $\vec{p} = p \left(\cos\phi\sin\theta, \sin\phi\sin\theta, \cos\theta\right)$, and introduce
$x=\cos\theta$ and $w=p/(2q)$. The momenta $k$ and $k'$ are then simply
\begin{align}
k = q\sqrt{1+2wx+w^2}, \qquad k' = q\sqrt{1-2wx+w^2}
\,.
\end{align}
Furthermore,
\begin{equation}
3\cos^2\widetilde{\theta} - 1 = 2 + \frac{3(x^2-1)}{1-2wx+w^2}
\,.
\end{equation}
The IR limit, $k=0$, is then given by the point $w=1$ and $x=-1$ in the $(w-x)$-plane. The other
momentum takes the value $k'=p=2q$ there. So, we can split the propagator as follows
\begin{align} \label{PropagatorSplitIntoFiniteAndIRpart}
&i\delta G_{+-}(\vec{y},\vec{q},\eta,\eta,k_0) \\
&\ = \frac{32\pi\mu^3}{3H_0^2}\int\!\frac{d^3\vec{p}}{(2\pi)^3}\Big[k'^2(3\cos^2\widetilde{\theta}-1) \mathrm{Im}\, J_{+-,-+}(\eta_0;\vec{p},\vec{q},\eta,\eta) {\rm e}^{i\vec{p}\cdot \vec{y}}- F^\mathrm{div}(\eta,\eta_0,\vec{p},\vec{q},\vec{y})\Big] \nonumber \\
&\quad + \frac{32\pi\mu^3}{3H_0^2}\int_{k\geq k_0} \frac{d^3\vec{p}}{(2\pi)^3}F^\mathrm{div}(\eta,\eta_0,\vec{p},\vec{q},\vec{y})
\,, \nonumber
\end{align}
with
\begin{eqnarray}
F^\mathrm{div}(\eta,\eta_0,\vec{p},\vec{q},\vec{y})
  &=& \frac{H_0^4 {\rm e}^{-2i\vec{q}\cdot\vec{y}}}{2\mu^2 p^2 k^3}
     \Bigg\{\frac{\sin(2q\eta_0)}{2q}\left(\cos(2\mu q\eta_0)
          -\frac{\sin(2\mu q\eta_0)}{2\mu q\eta_0}\right)
\\
  && -\, \mu^2 \bigg[\eta_0 \cos(2q\eta_0)
           \frac{\sin(2\mu q\eta_0)}{2\mu q\eta_0}
       - \frac{\sin(2q\eta_0)}{2q} \cos(2\mu q\eta_0)\bigg] \Bigg\}
\,.
\nonumber
\end{eqnarray}
Restricting the momentum $k$ to be above some IR cut-off $k_0$ translates to
\begin{equation}
x\geq x_0\left(w,\frac{k_0}{q}\right)= \frac{1}{2w}\left(\frac{k_0^2}{q^2}-(1+w^2)\right)
\end{equation}
for $1-k_0/q\leq w\leq 1+k_0/q$. It follows that
\begin{equation}
\int_{k\geq k_0} \frac{dw dx}{(1+2wx+w^2)^{3/2}}=\log\left(\frac{q^2}{k_0^2}\right) + 2 + \mathcal{O}(k_0).
\end{equation}
We can neglect terms that vanish for $k_0 \rightarrow 0$. As a result, we obtain an explicit
expression for the cut-off dependent part of the propagator
\begin{align}
&\frac{32\pi\mu^3}{3H_0^2}\int_{k\geq k_0}\frac{d^3\vec{p}}{(2\pi)^3}F^\mathrm{div}(\eta,\eta_0,\vec{p},\vec{q},\vec{y}) \\
&\qquad = \frac{8\mu H_0^2 {\rm e}^{-2i\vec{q}\cdot\vec{y}}}{3\pi q^2}\Bigg\{\frac{\sin(2q\eta_0)}{2q}\left(\cos(2\mu q\eta_0)-\frac{\sin(2\mu q\eta_0)}{2\mu q\eta_0}\right) \nonumber \\
&\qquad\quad - \mu^2 \bigg[\eta_0 \cos(2q\eta_0)\frac{\sin(2\mu q\eta_0)}{2\mu q\eta_0} - \frac{\sin(2q\eta_0)}{2q} \cos(2\mu q\eta_0)\bigg] \Bigg\}\left(\log\left(\frac{q^2}{k_0^2}\right) + 2\right).
\nonumber
\end{align}
 The power spectrum is defined by ({\it cf.} Eq.~(\ref{power spectrum:dS})),
\begin{equation}
\langle \Omega |\phi(\vec{x},\eta)\phi(\vec{x}',\eta)|\Omega\rangle
 = \int \frac{dq}{q} \mathcal{P}_\phi(\vec q,\vec{y},\eta)\frac{\sin(qr)}{qr}
\,,
\end{equation}
with $r=\|\vec{x}-\vec{x}'\|$. We have already derived the power spectrum for the de~Sitter
background, equation~(\ref{dSSpectrum}). Hence, we will only be interested in the correction
induced by the black hole,
\begin{equation} \label{PropagatorToSpectrum}
\delta \mathcal{P}_\phi(\vec q,\vec y,\eta)
 = \frac{q^3}{2\pi^2}\ \mathrm{Re}\, i\delta G_{+-}(\vec{y},\vec{q},\eta,\eta)\,.
\end{equation}
Again, we can make a split into IR finite and IR divergent part,
\begin{equation} \label{SplitSpectrumIntoFiniteAndIRpart}
\delta\mathcal{P}_\phi(\vec q,\vec y,\eta)
  = \delta\mathcal{P}_\phi^{\mathrm{fin}}(\vec q,\vec y,\eta)
      + \delta\mathcal{P}_\phi^{\mathrm{div}}(\vec q,\vec y,\eta,k_0),
\end{equation}
with
\begin{eqnarray}
\delta\mathcal{P}_\phi^{\mathrm{div}}(\vec q,\vec y,\eta,k_0)
  = -\frac{4\mu q H_0\cos(2\vec q\cdot\vec y)}{3\pi^3}
   \Bigg\{\frac{\sin(2q/H_0)}{2q/H_0}\left(\cos(2\mu q/H_0)
           -\frac{\sin(2\mu q/H_0)}{2\mu q/H_0}\right)
\nonumber \\
  -\, \mu^2 \bigg[\cos(2q/H_0)\frac{\sin(2\mu q/H_0)}{2\mu q/H_0}
 - \frac{\sin(2q/H_0)}{2q/H_0} \cos(2\mu q/H_0)\bigg]
   \Bigg\}\left(\log\left(\frac{q^2}{k_0^2}\right) + 2\right)
\,.
\nonumber
\end{eqnarray}
This expression will be added to the numerical results for the finite part to determine the total
correction. The finite part is given in terms of an integral in Appendix C,
Eq.~(\ref{CorrectionFullIntegralExpression}).

\section{Numerical results}
\label{Numerical results}

\subsection{The anisotropic case: different angles}
\label{The anisotropic case: different angles}

\begin{figure}
\centering
\includegraphics[scale=0.6]{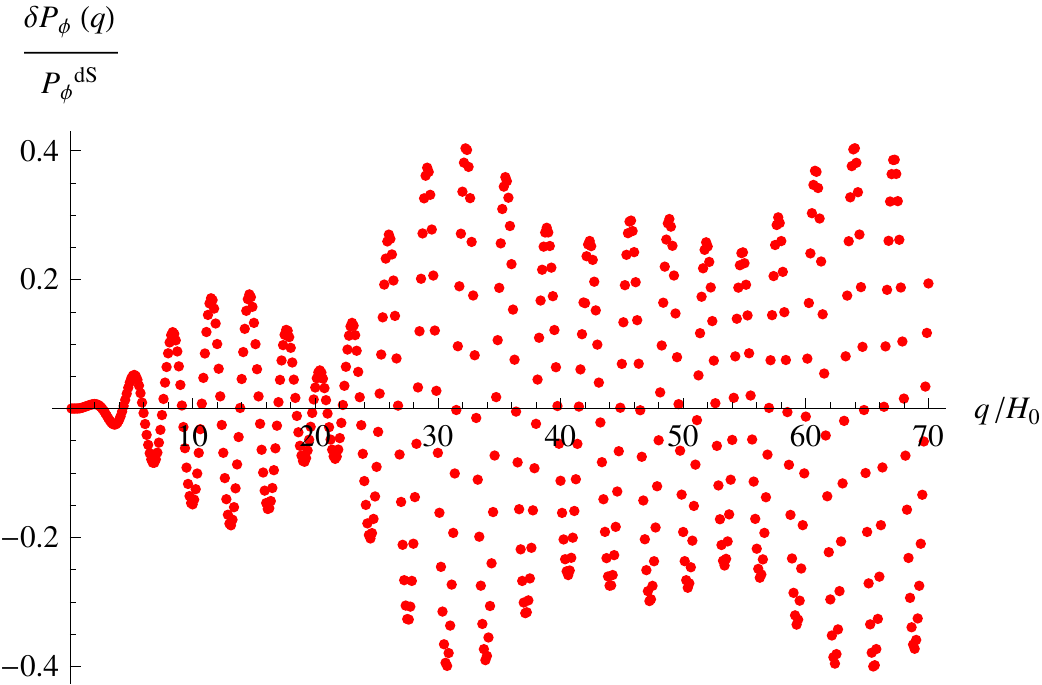}
\includegraphics[scale=0.6]{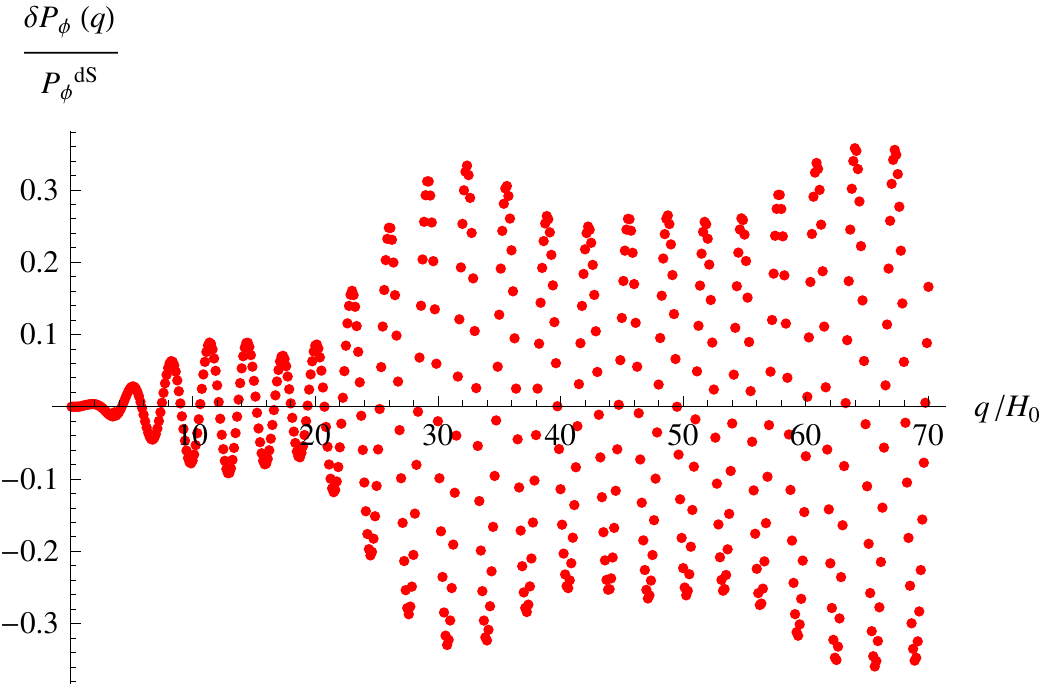}
\includegraphics[scale=0.6]{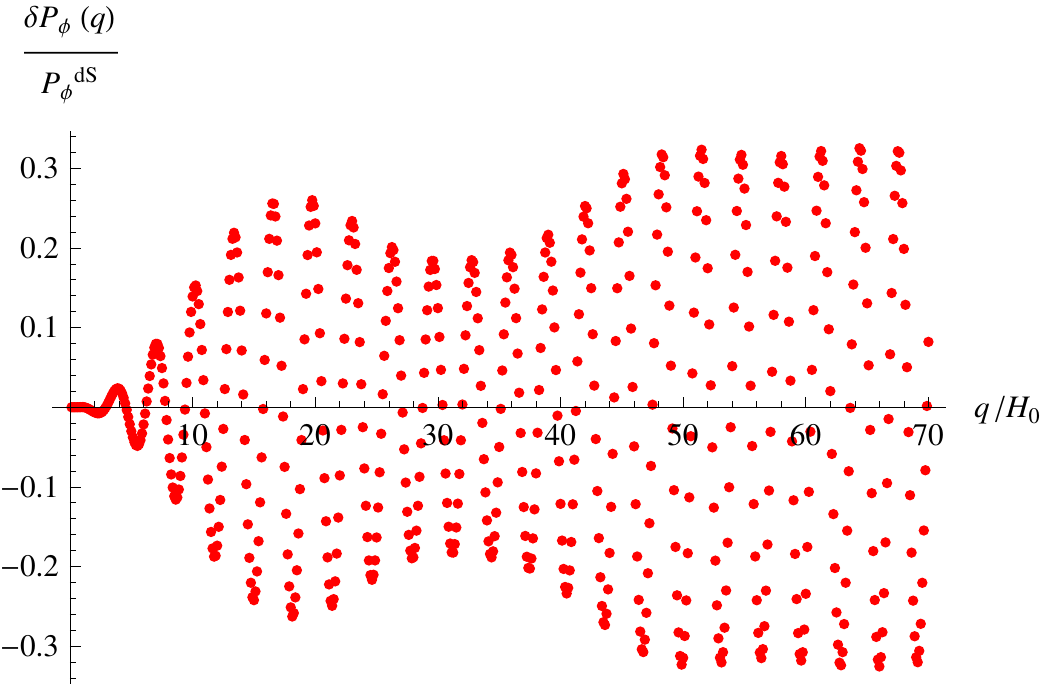}
\includegraphics[scale=0.6]{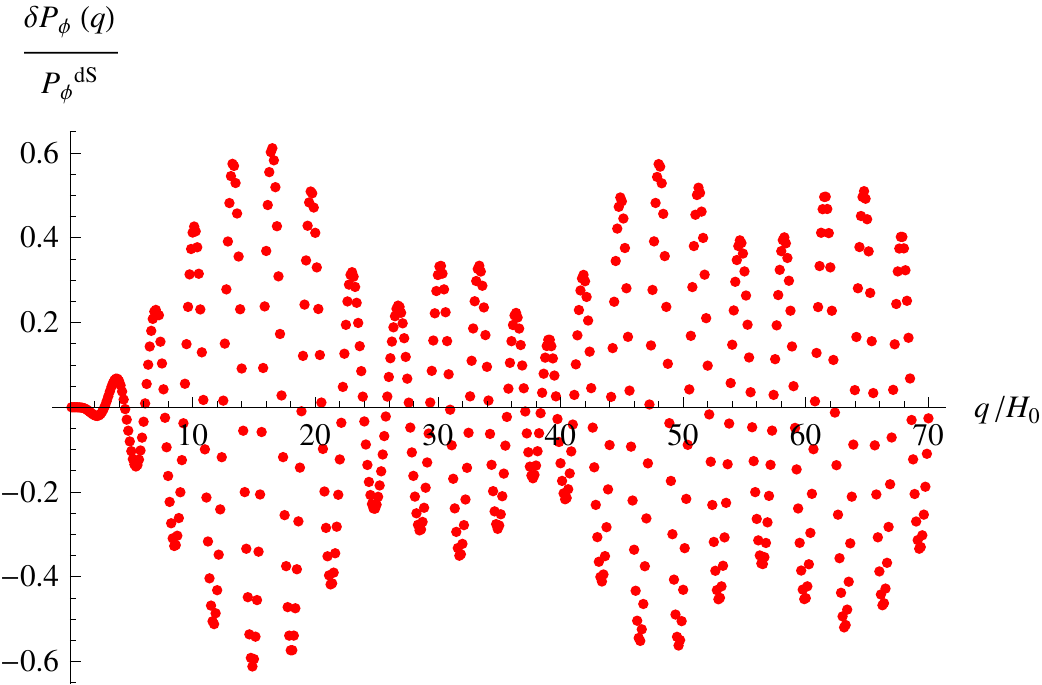}
\caption{\footnotesize The correction to the spectrum is plotted as a function of the momentum $q$
for fixed $qy=1$, $\mu=1/10$ and cut-off $k_0/H_0=1/10$ but different angles between $\vec q$ and
$\vec y$. Upper left panel: $\theta=0$, upper right: $\theta=\pi/8$, lower left: $\theta=\pi/4$ and
lower right: $\theta=\pi/4$. The case $\theta =\pi/2$ is treated separately (Fig.~\ref{fig:
y0vsThetaPiOver2}). This is the result of a numerical integration with a discretization of eight
points per unit. For general angles we could not find a good fitting curve. Nevertheless, we
observe $\cos(2q/H_0)$ oscillations modulated with $\cos(2\mu q/H_0)$.} \label{fig: qy1Spectrum}
\end{figure}

\begin{figure}
\centering
\includegraphics[scale=0.9]{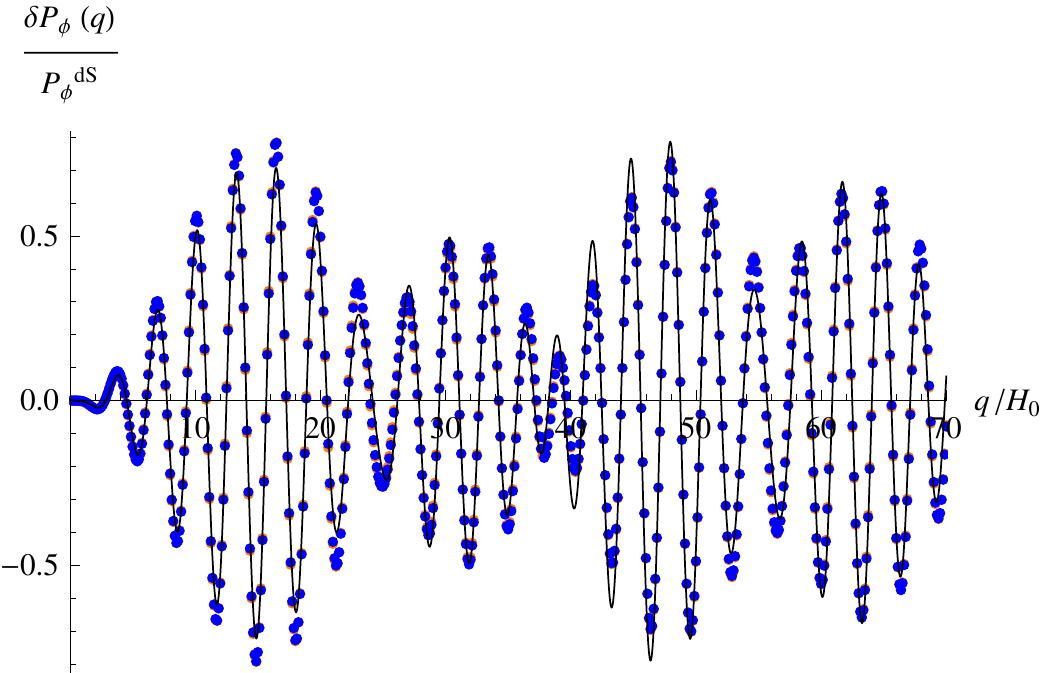}
\caption{\footnotesize It is shown that the anisotropic case (blue dots, dark) with vectors
$\vec{q}$ and $\vec{y}$ being perpendicular, $\theta=\pi/2$, coincides with the isotropic case,
$\vec{y}=0$ (orange dots, light, mostly covered by blue dots). This is illustrated for mass
parameter $\mu=1/10$, cut-off $k_0/H_0=1/10$ and $qy=1$. The black fitting curve is given
explicitly in Appendix D.} \label{fig: y0vsThetaPiOver2}
\end{figure}

\begin{figure}
\centering
\includegraphics[scale=0.6]{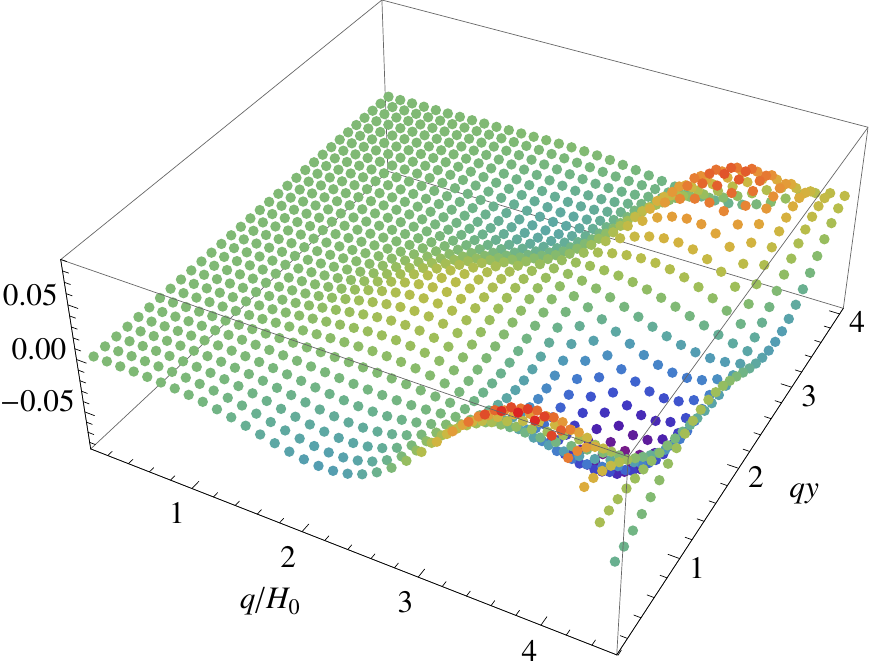}
\includegraphics[scale=0.6]{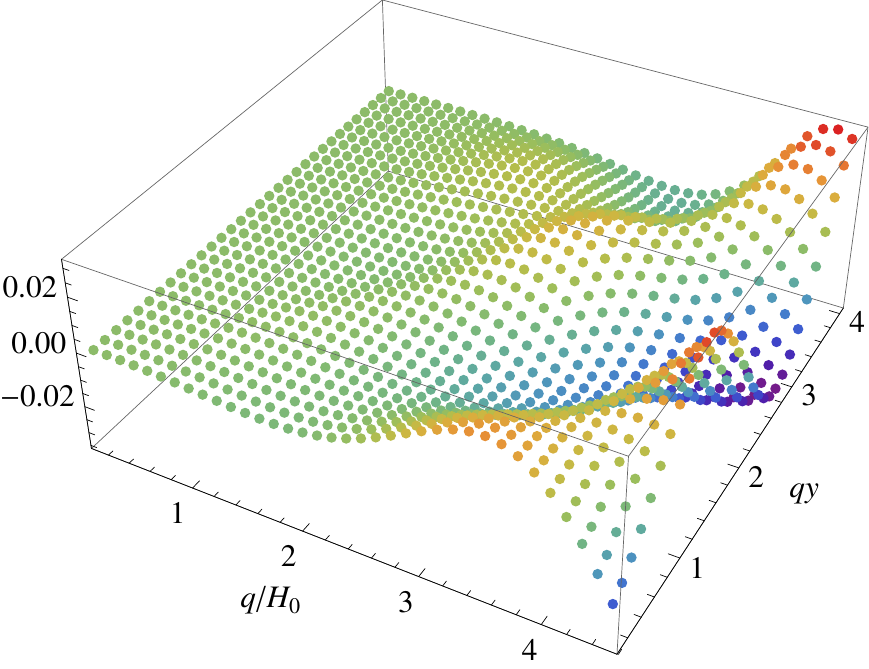}
\caption{\footnotesize This plot shows the numerical result for the spectrum for small values of
the two variables $q$ and $qy$ for $\mu=1/10$ and $\sphericalangle(\vec q,\vec y)=0$. A
discretization of eight points per unit was used. The momentum cut-off is $k_0/H_0=1/10$ for the
left panel and $k_0/H_0=1$ for the right one. There is a significant cut-off dependence.}
\label{fig: SpectrumSmallValues}
\end{figure}

For $\vec{y}\neq 0$ we can study the dependence of the spectrum on the angle between $\vec{q}$ and
$\vec{y}$. The numerical results are shown in Fig.~\ref{fig: qy1Spectrum}. All plots show the
correction to the spectrum $\delta \mathcal{P}_\phi$ normalized by the scale invariant de~Sitter
spectrum, Eq.~(\ref{dSSpectrum}),
\begin{equation}
\mathcal{P}_\phi^{\mathrm{dS}}\Big|_{\eta=0}=\frac{H_0^2}{4\pi^2}\,.
\end{equation}
Although a good fitting curve has not been found for general angles, it is worth noting that
besides the high frequency oscillations a modulation with a much lower frequency is present which
is determined by the mass parameter $\mu$. Therefore, the mass of the black hole can, in principle,
be inferred directly from the corrections to the scale invariant de~Sitter spectrum at large $q$,
\textit{i.e.} at small scales. However, the modes of this magnitude that are stretched into today's
observable scales are due to black holes that were created when the Hubble radius was much smaller
than today's Hubble radius. The probability to find such black holes inside our Hubble volume was
estimated in section~\ref{Formation probability for black holes}. We will come back to this point
in the discussion (see also Fig.~\ref{fig: HorizonCrossing}).
\paragraph{}
The anisotropic case, $\vec{y} \neq 0$, with the vectors $\vec y$ and $\vec q$ being perpendicular
($\theta=\pi/2$) turns out to be identical within numerical precision to the isotropic case,
$\vec{y}=0$. From the point of view of observations it is hence impossible to associate data to the
one case or the other. On the level of the integral
expressions~(\ref{CorrectionFullIntegralExpression}, Appendix C) for the two cases one can argue
that, because of the pole at $(w,x)=(1,-1)$, the Bessel function $J_0(2qyw\sqrt{1-x^2})$ that is
present in the anisotropic case is effectively evaluated to unity, yielding the same result as the
isotropic case. We demonstrate this in Fig.~\ref{fig: y0vsThetaPiOver2}.
Small values of $q$ (large scales) are relevant for black holes that formed at a later stage of
inflation although their formation probability is exponentially suppressed (see discussion
section). We present numerical results for the spectrum in the region of small $qy$ and $q$ in
Fig.~\ref{fig: SpectrumSmallValues}.
Furthermore, for $q/H_0 \ll 1$ we can expand $\delta \mathcal{P}_\phi(\vec q,\vec y)$ and find that
\begin{equation} \label{Factorization}
\delta \mathcal{P}_\phi(\vec q,\vec y,k_0) = \frac{16 \mu^5 q^3}{9\pi^3H_0}F(qy, \theta) + \frac{16 \mu^5 q^3}{9\pi^3H_0}\left(\log\left(\frac{q^2}{k_0^2}\right)+2\right)\cos(2qy\cos\theta),
\end{equation}
with $\theta=\sphericalangle(\vec q,\vec y)$. This means that the cut-off dependent and cut-off
independent part of the correction to the spectrum factorize individually for low momenta. The
integral expression of the functions $F$ is given in Appendix~C,
Eq.~(\ref{FactorizationFunctionF}). In Fig.~\ref{fig: Factorisation} we plot the corrected spectrum
for different angles and small $q/H_0$. We find that the function $F$ is very well approximated by
\begin{equation} \label{FactorisationFit}
F(qy,\theta) =
\begin{cases}
-4.8 (\log(qy) + 1.2) \cos\left(2 qy \cos(\theta)\right) & \text{for } qy \lesssim 0.5, \\
-2.0 (\log(qy) + 1.3) \cos\left(2 qy \cos(\theta)\right) & \text{for } qy \gtrsim 2.0.
\end{cases}
\end{equation}
for any angle $\theta$. The fit is not very good in the intermediate region $0.5 < qy < 2.0$. Thus,
we have found an analytic expression that describes the correction to the spectrum very well for
large scales. Moreover, the mass parameter $\mu$ for the black hole can be found from the amplitude
of the spectrum and the IR cut-off $k_0$ from the enveloping curve.

\subsection{The isotropic case: cut-off and $\mu$ dependence}

We pointed out in the previous section that the isotropic case virtually coincides with the case
that the vectors $\vec q$ and $\vec y$ are perpendicular but non-zero. Yet, we can study the
dependence of the spectrum on $\mu$ and the cut-off $k_0$. The results are summarized in
Fig.~\ref{fig: SpectrumIsotropic} with general fitting curves (\textit{cf.} Appendix D) that
approximate the numerical data very well, in particular for small $\mu$. Again, the mass parameter
determines the modulation frequency.

\section{Discussion and Conclusion}
\label{Discussion and Conclusion}

 In this work we have derived the correction to the scale invariant power spectrum of a scalar field
on de~Sitter space from a small primordial black hole to lowest orders in its mass parameter
$\mu=(GMH/2)^{1/3}$. To this end, in section~\ref{Formation probability for black holes}, we have
first analyzed the probability of black hole formation in the pre-inflationary Universe. In order
to maximize the formation probability, we have assumed that the pre-inflationary Universe is
dominated by heavy non-relativistic particles, in which case sub-Hubble perturbations grow. We have
found that there is a range for the particle mass $m<m_p$
\begin{figure}[h]
\centering
\includegraphics[scale=0.6]{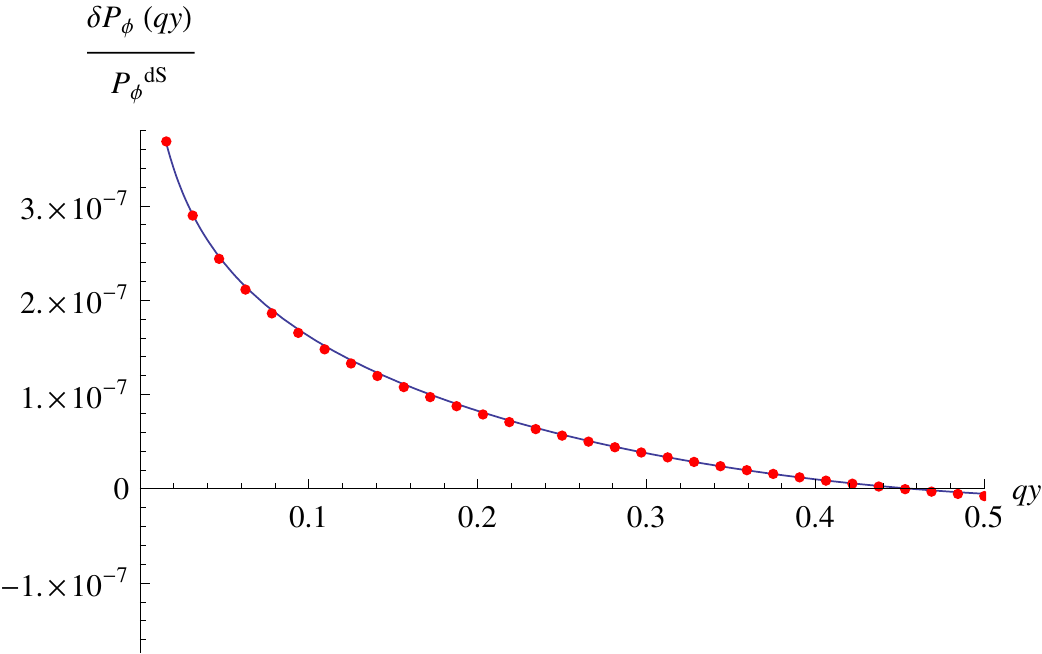}
\includegraphics[scale=0.6]{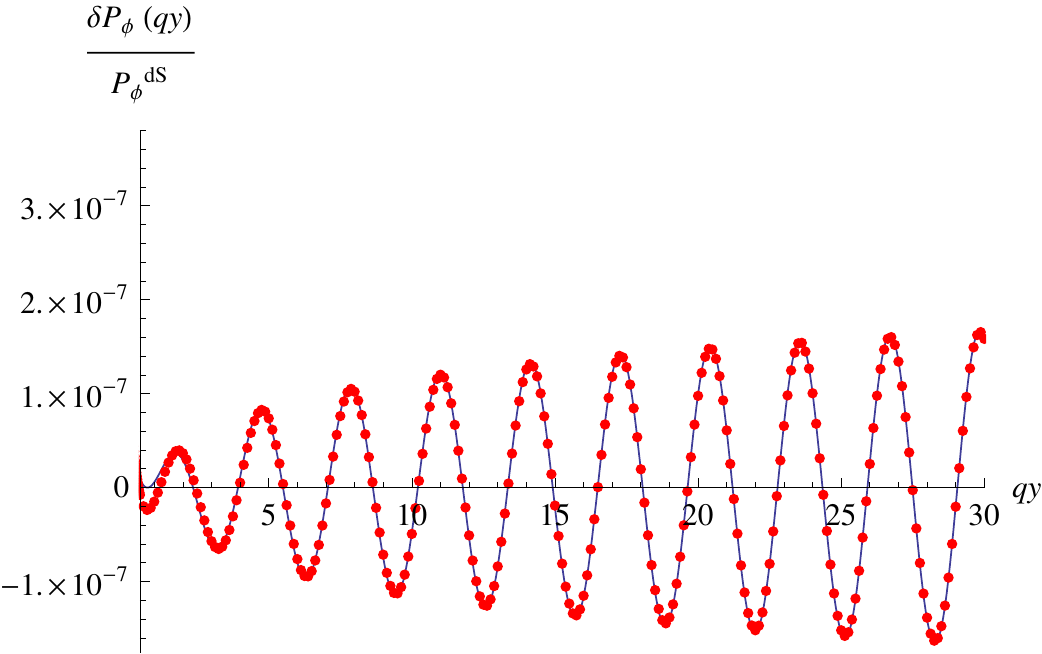}
\includegraphics[scale=0.6]{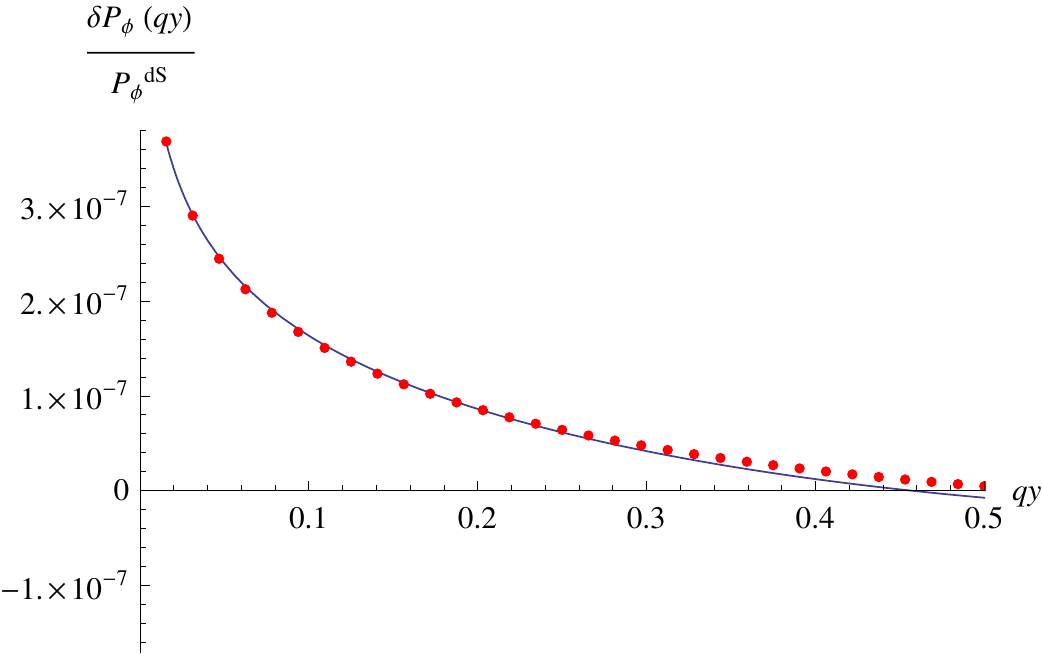}
\includegraphics[scale=0.6]{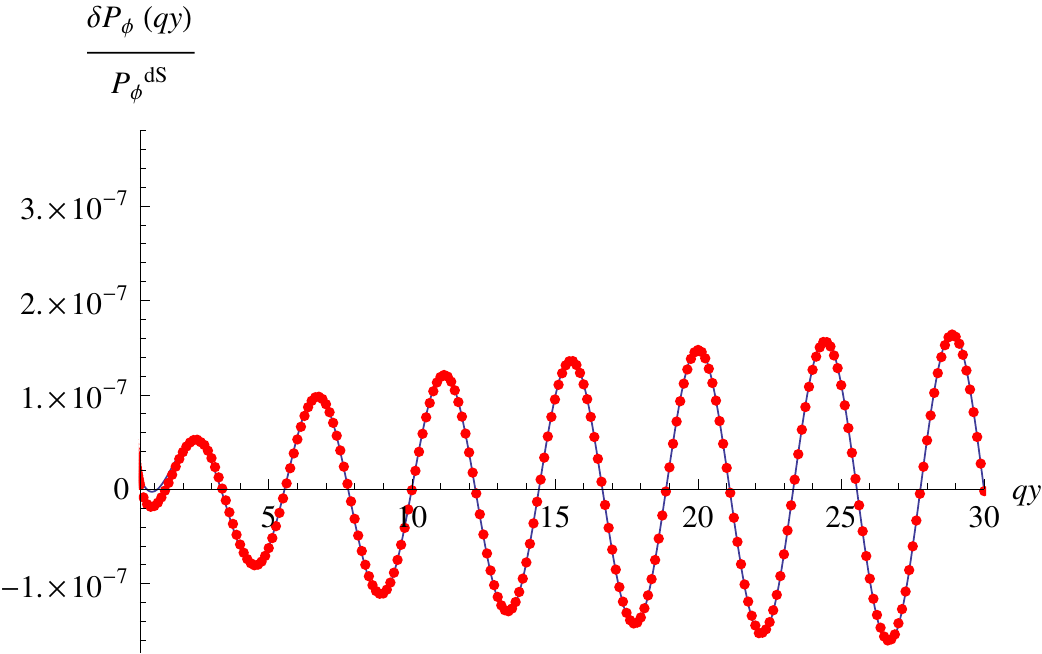}
\includegraphics[scale=0.6]{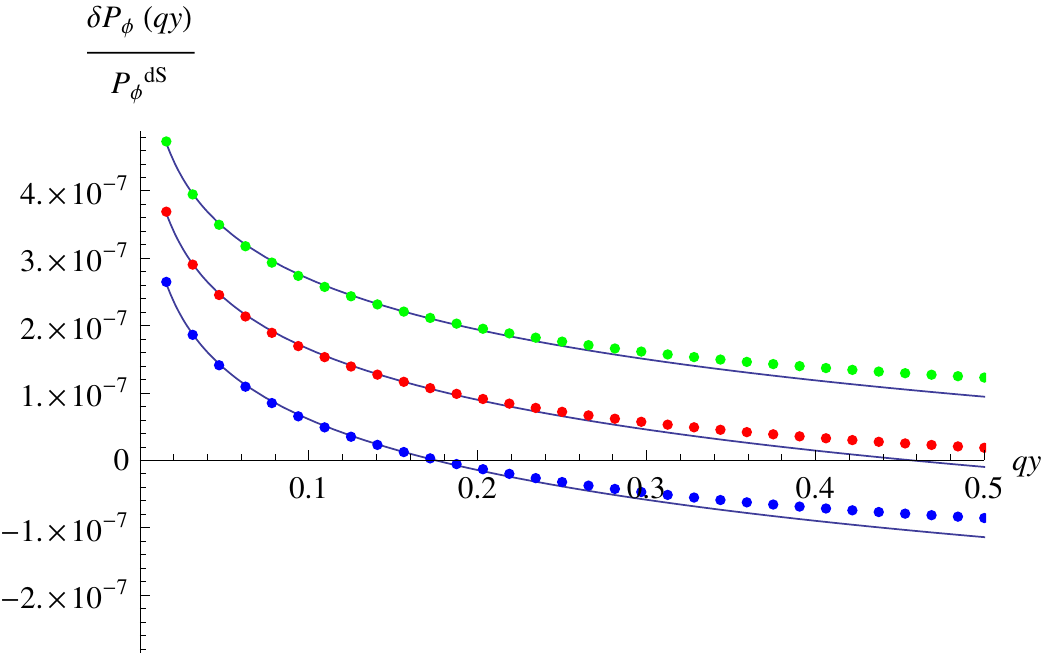}
\includegraphics[scale=0.6]{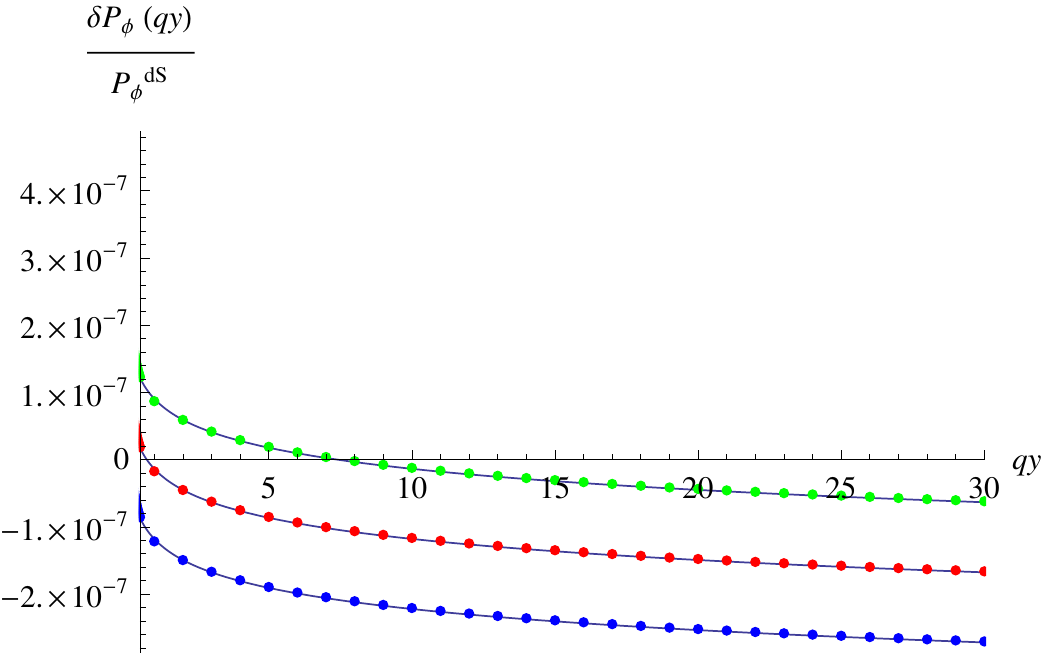}
\caption{\footnotesize The correction to the spectrum as a function of $qy$ is plotted for
different angles $\theta$ between $\vec{q}$ and $\vec{y}$ in the limit $q/H_0\rightarrow 0$. Upper
panels: $\theta=0$, middle panels: $\theta=\pi/4$ and lower panels: $\theta=\pi/2$. For all plots
we chose $q/H_0 =1/10$ and $\mu=1/10$. The upper and middle panels have a cut-off $k_0/H_0=1/10$.
The lower panels show also the cut-off dependence: green (upper) dots correspond to $k_0/H_0 =
1/100$, red (middle) dots to $k_0/H_0=1/10$ and blue (lower) dots to $k_0/H_0=1$. The fitting
curves are given by Eq.~(\ref{FactorisationFit})} \label{fig: Factorisation}
\end{figure}
and the Hubble rate $H_*<m_p$ during that period for which the expected number of sub-Hubble black holes per Hubble volume can be larger than
$\mathcal{O}(1)$, as can be seen in Fig.~\ref{fig: FormationProbability}.

 To determine the correction to the spectrum in section~\ref{PropagatorSection} we have derived an
analytic expression for the momentum space propagator of the massless, minimally coupled, scalar
field on the Schwarzschild-de~Sitter background in the Schwinger-Keldysh formalism. We observe that
the propagator diverges in the infrared, demonstrating that, in contrast to a recent proposal
\cite{Carroll:2008br}, an expansion in the momenta is inappropriate, although the breaking of
homogeneity is weak. This divergence can be traced to the well known infrared divergence of the
massless scalar propagator on de Sitter. We have used a simple regularization, which consists of
placing the Universe in a large, but finite, comoving box.
\begin{figure}[h]
\centering
\includegraphics[scale=0.6]{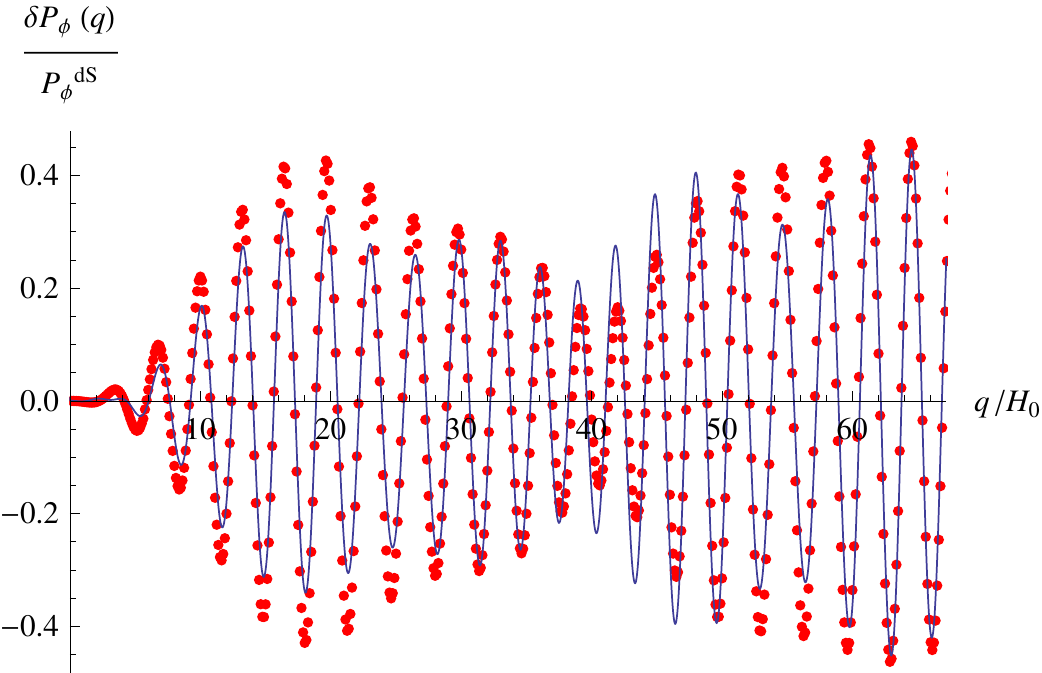}
\includegraphics[scale=0.6]{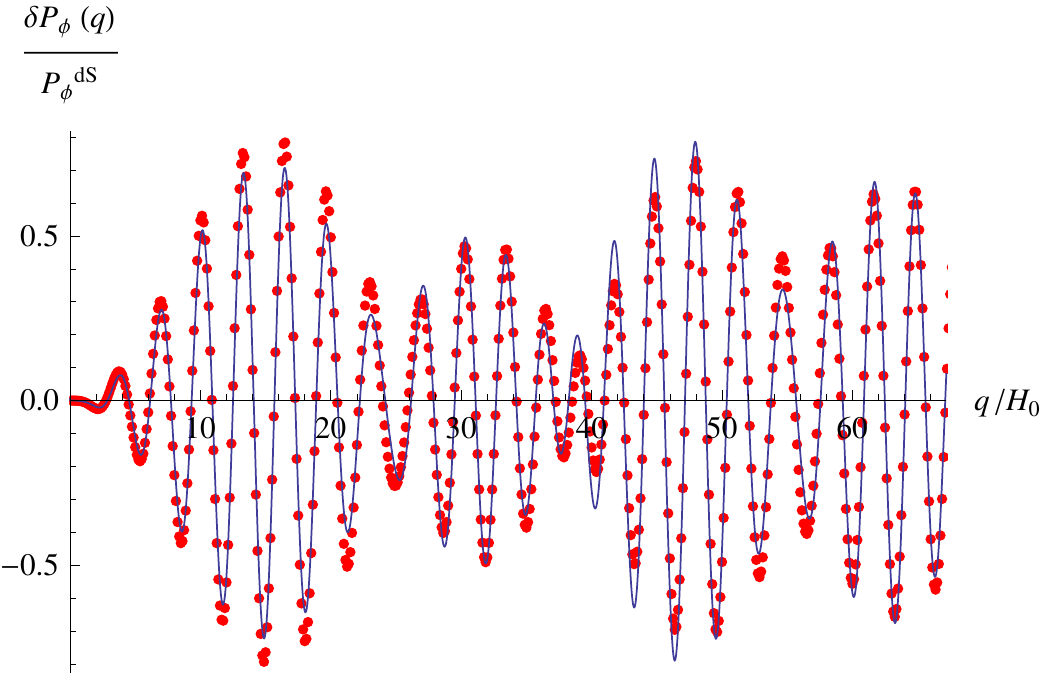}
\includegraphics[scale=0.6]{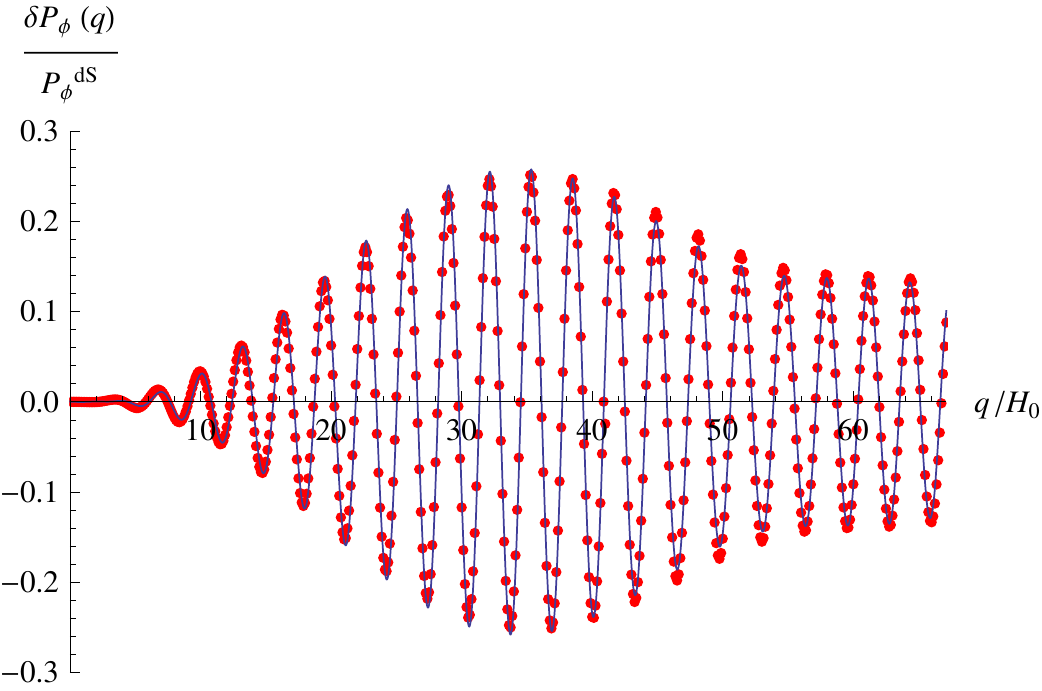}
\includegraphics[scale=0.6]{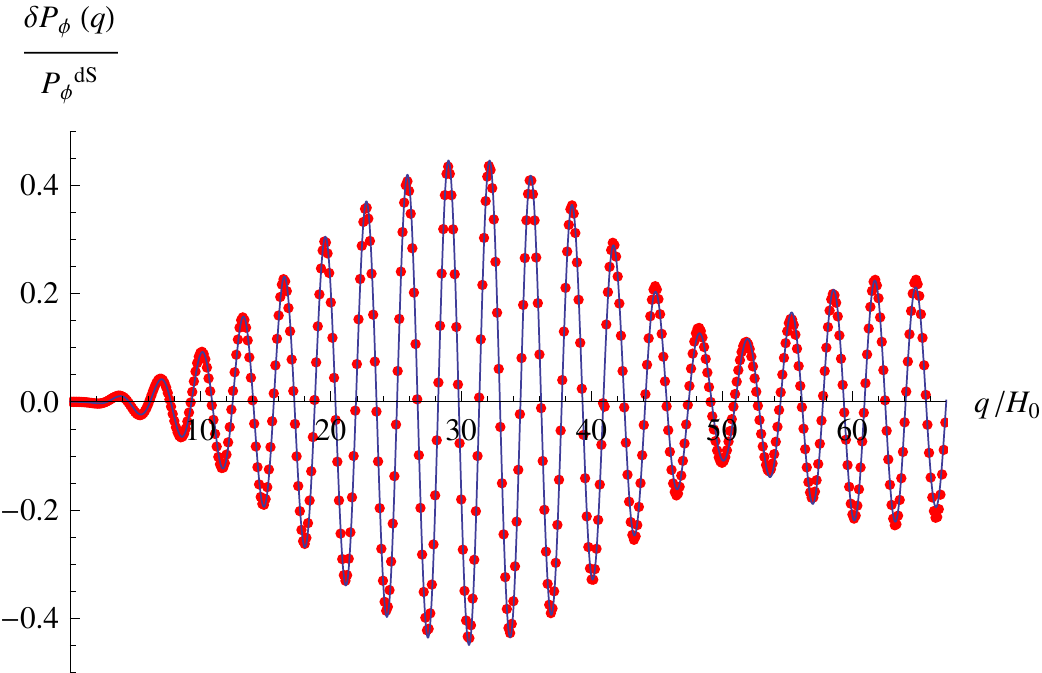}
\includegraphics[scale=0.6]{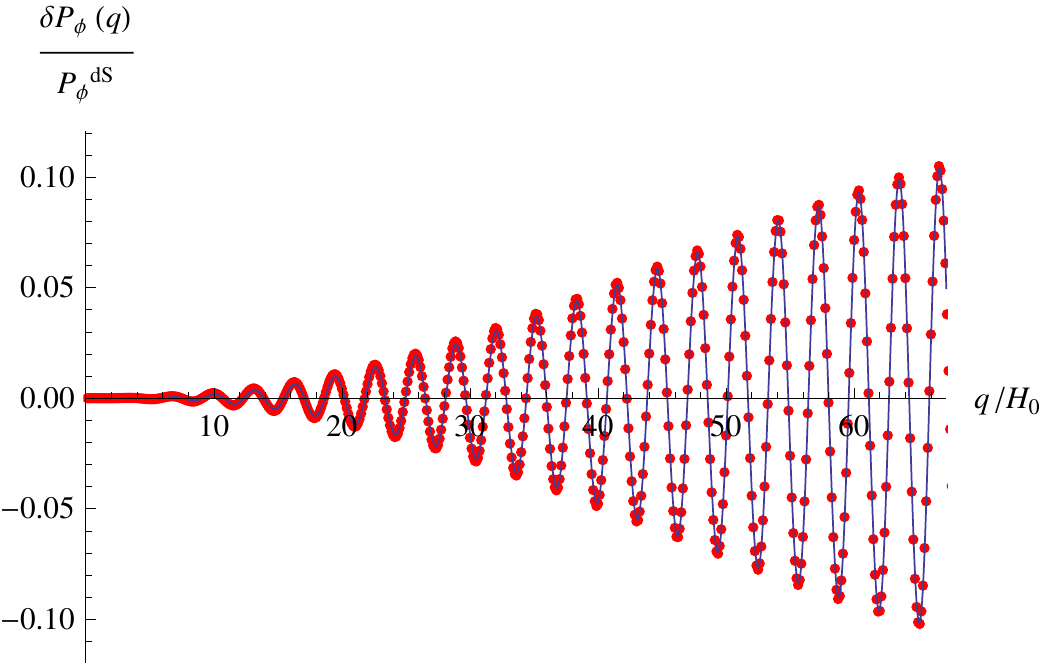}
\includegraphics[scale=0.6]{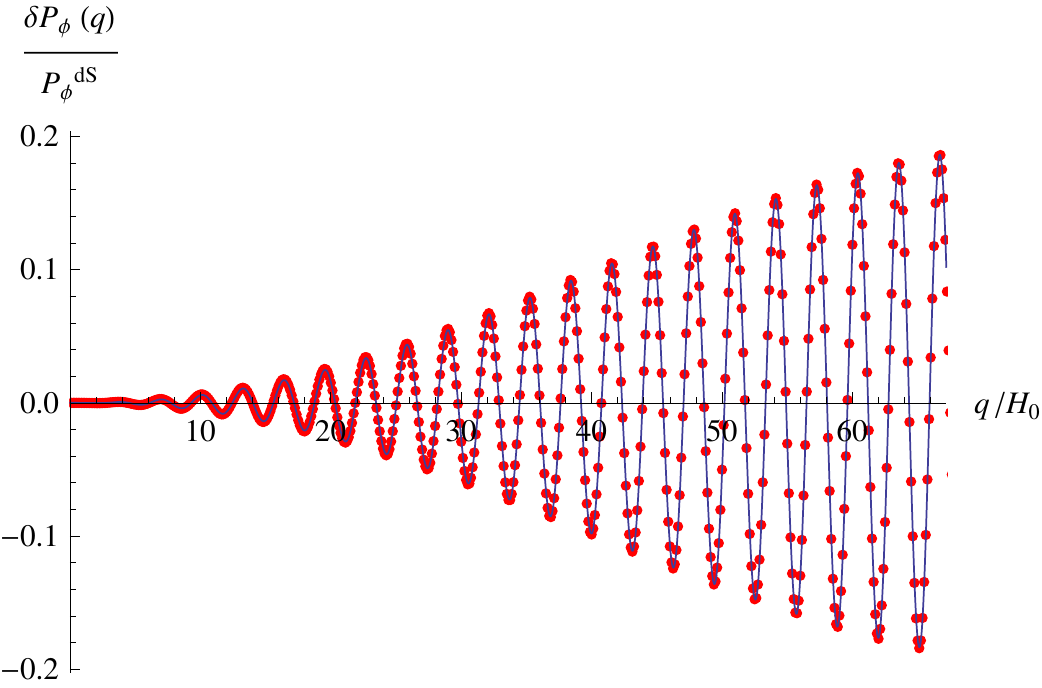}
\caption{\footnotesize We present here the numerical data (red dots) for the correction to the
spectrum normalized by the de~Sitter spectrum in the isotropic case for cut-off $k_0/H_0=1$ (left
panels) and $k_0/H_0=1/10$ (right panels). Upper panels: $\mu=1/10$, middle: $\mu=1/20$ and lower:
$\mu=1/50$. The analytic expression for the fitting function (blue curve) can be found in Appendix
D.} \label{fig: SpectrumIsotropic}
\end{figure}
In section~\ref{CMBsection} we have outlined the procedure which has allowed to determine the impact of an inflationary black holes on
the CMB and structure formation. We have demonstrated that, to leading order in the dimensionless
black hole mass parameter $\mu=(GMH/2)^{1/3}$, the Sasaki-Mukhanov field (curvature perturbation)
remains the correct gauge invariant, dynamical scalar perturbation. By working in the zero
curvature gauge, we have then shown how to connect the spectrum in the scalar field fluctuation to
the spectrum of the comoving curvature perturbation. Finally, the knowledge of the appropriate
transfer functions has allowed to relate the inflationary curvature perturbation to the CMB
temperature fluctuations and to the large scale structure of the Universe. That analysis is yet to
be done in detail. Temperature fluctuations are obtained from fluctuations in the Sasaki-Mukhanov
variable by the Sachs-Wolfe effect which yields on large scales
\begin{equation} \label{TempFlucSMvarPositionSpace}
\bigg\langle \frac{\delta T(\vec{x},t) \delta T(\vec{x}',t) }{T_0^2}\bigg\rangle =
\frac{1}{9}\langle \Omega |\mathcal{R}(\vec{x},t)\mathcal{R}(\vec{x}',t) |\Omega \rangle\,.
\end{equation}
The power spectrum $\mathcal{P_R}$ of scalar cosmological perturbations was studied in the mixed
representation in the sections~\ref{MixSpaceRepSection} and~\ref{Numerical results} and can thus be
related to the temperature fluctuations on large scales by performing a Fourier transformation
of~\eqref{TempFlucSMvarPositionSpace} with respect to $\vec{r}=\vec{x}-\vec{x}'$. On small scales
the more complicated relation~\eqref{temperature corr fn} has to be used, whereby the appropriate
transfer functions have to be determined numerically. Other transfer functions have to be used if
one wants to study the effect on large scale structure from perturbations of $\mathcal{R}$.

The scalar field propagator in the mixed representation is closely related to the Wigner function,
and hence admits a probabilistic interpretation characterizing the Boltzmann distribution function.
In section~\ref{Numerical results} we devoted quite some effort to analyze the mixed space spectrum
which is a function of not only the relative (comoving) momentum $q=\|\vec q\|$, but also of the
comoving black hole distance from us, $y=\|\vec y\|$, of the angle $\sphericalangle (\vec q,\vec
y)$, and finally of the lowest infrared (cut-off) momentum $k_0$ that can be excited. Our results
are mostly analyzed as the black hole contribution to the spectrum relative to the scalar
contribution in de Sitter space. Since the observed spectrum is highly isotropic, and seemingly
homogeneous, we were primarily interested in the case when the perturbation induced by a black hole
is small, which led us to consider the limit $\mu\ll 1$, or, more precisely, $0.027<\mu\ll 1$, {\it
cf.} Eq.~(\ref{mu:limits}). The lower limit comes from the requirement that, before it evaporates,
the black hole must last at least several e-folds during inflation. The effect of the black hole
evaporation during inflation is illustrated in Fig.~\ref{fig: Mu_efolding}.

 The spectrum is first analyzed in section~\ref{The anisotropic case: different angles} for the
general anisotropic case, $\vec y \neq 0$, with $\vec y$ the displacement vector of the black hole
with respect to us for different angles between $\vec y$ and the momentum vector $\vec q$ that is
conjugate to the relative distance of two points. Then we considered the large scale region
($q/H_0$ and $qy$ small) for the special case that $\vec q$ and $\vec y$ are parallel. Furthermore,
by making an expansion for $q \ll H_0$ of the integral expression for the spectrum we showed that
it takes a relatively simple form, Eq.~(\ref{Factorization}). We presented an explicit expression
for fitting functions for the $qy$ dependent part which was determined numerically. In the
isotropic case, $\vec y=0$, the dependence on the parameter $\mu$ is shown. Unlike the spectrum of
scalar homogeneous perturbations in inflation, which is a function of the momentum magnitude $q$,
and depends on two parameters, $H$ and $\epsilon=-\dot H/H^2$ at the Hubble crossing, the scalar
spectrum in Schwarzschild-de Sitter space depends on $q$, the distance to the black hole $y$ and
the angle between $\vec q$ and $\vec y$. It can serve as a six dimensional template, whereby the
template parameters are the comoving black hole position $\vec y$, its mass parameter $\mu$, and
$H$ and $\epsilon=-\dot H/H^2$ at the first Hubble crossing during inflation. As a summary of the
numerical results we can conclude that (i) the spectrum as a function of $q$ is modulated with a
lower frequency which is characterized by the mass parameter $\mu$ of the black hole; (ii) the
enveloping amplitude of the spectrum scales logarithmically with the IR cut-off $k_0$, and (iii)
the isotropic case cannot be distinguished from a particular configuration of the anisotropic case
(where $\vec q\perp \vec y)$.

 We should point out, however, that a comprehensive analysis of the power spectrum of scalar
cosmological perturbations induced by small inflationary black holes should take into account also
the degrees of freedom of the graviton and start with the action for the scalar field and the
graviton on the unperturbed background. In particular, it is important to study the possible mixing
of the scalar, vector and tensor modes induced by the black hole. Provided
Eq.~(\ref{spectrum:comoving gauge}) is an accurate expression for the curvature perturbation in
terms of the scalar field perturbation in the inhomogeneous case when a small black hole is
present, then based on Eq.~(\ref{spectrum:comoving gauge}) one can calculate the spectrum of
comoving curvature perturbation induced by an inflationary black hole. This is the quantity of
interest since it sources the CMB temperature fluctuations and the large scale structure. Of
course, a more realistic inflationary background is a Schwarzschild black hole in a quasi-de~Sitter
space (SqdS), in which the Hubble rate and the deceleration parameter are both slowly varying
functions of time. In order to make sure that the results presented here hold also for that case,
one would have to make a complete analysis of cosmological perturbations on a SqdS space, which is
beyond the scope of the present work.

\begin{figure}[h]
\centering
\includegraphics[scale=0.9]{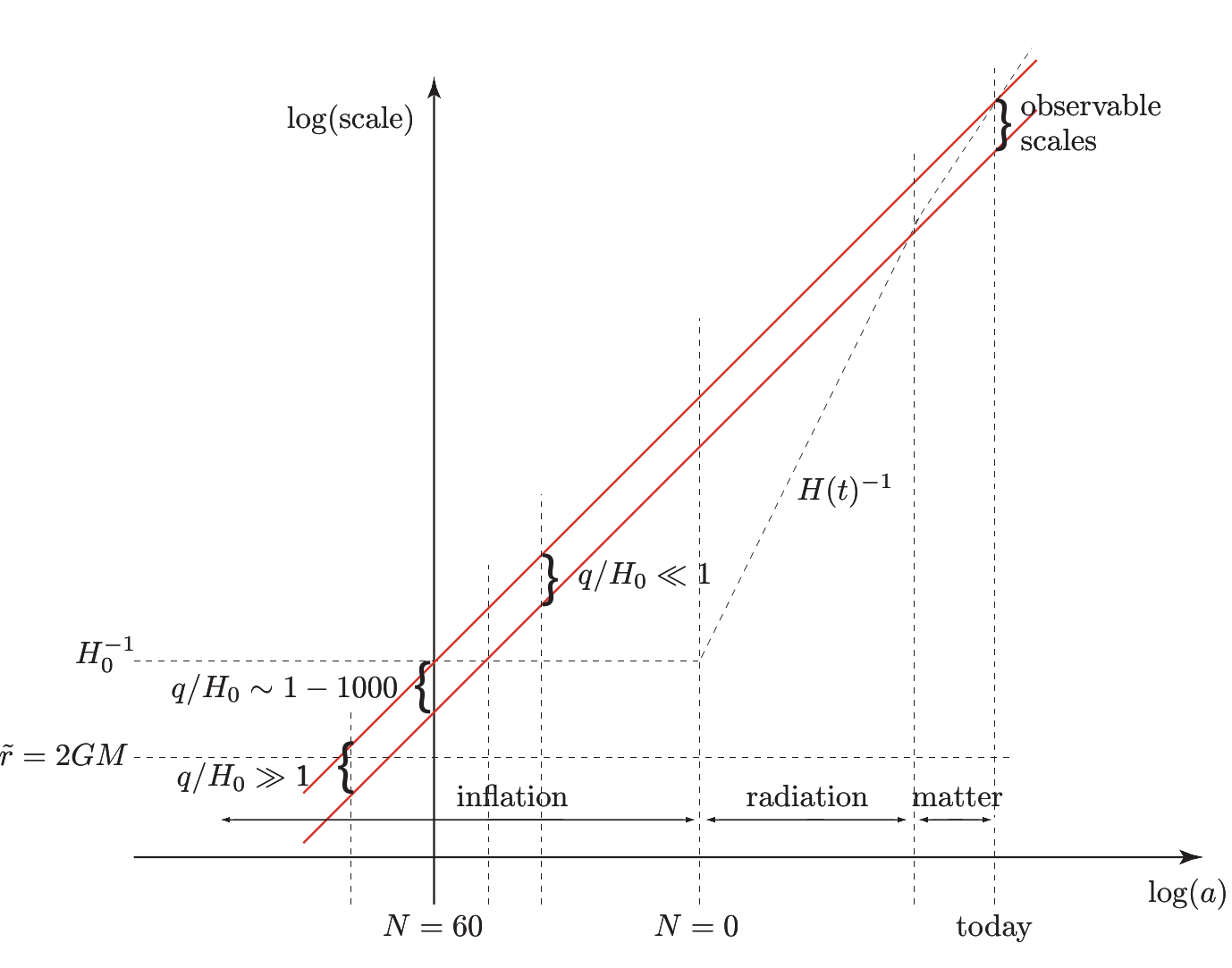}
\caption{\footnotesize The stretching of modes during the evolution of the Universe is displayed.
The time axis is $\log(a/a_{\rm end})=-N$ with $N$ the number of e-foldings before the end of
inflation. During matter domination the Hubble rate is given by $H(t)\propto a^{-3/2}$, during the
radiation era it is $H(t)\sim a^{-2}$. A constant Hubble rate $H_0$ during inflation (slow-roll
approximation) is assumed. The upper diagonal line (red) shows the stretching of the mode that
re-enters the Hubble volume today, thereby defining the largest observable scale. That mode crossed
the horizon approximately $N_0\simeq 60$ e-foldings before the end of inflation (sometimes a value
of $N_0=50$ is assumed). The smallest observable scales depend on the maximum resolution of the
apparatus used for the measurement. Today's best resolution is $l\sim \pi/\theta(\mathrm{rad})\sim
10^2$. The lower diagonal line (red) marks this mode, it exited the Hubble horizon at $N=50$.}
\label{fig: HorizonCrossing}
\end{figure}

 Corrections to the power spectrum from primordial black holes are potentially relevant for
observations. To assess this possibility more carefully, we have to establish a relationship
between the probability of pre-inflationary black hole formation from section~\ref{Formation
probability for black holes} and the expected number of black holes that leave an imprint on
today's CMB sky. A first step in doing that is to trace physical wavelengths,
$\lambda_\mathrm{ph}=a/q$, back in time and check which modes correspond to today's observable
scales. In Fig.~\ref{fig: HorizonCrossing} it is shown how physical scales are stretched during the
evolution of the Universe. The time axis is given in terms of the scale factor $\log(a)$. We are
interested in the modes which cross for the first time during inflation the Hubble scale at $t_{\rm
1x}$ when $q=a_{\rm 1x}H_{\rm 1x}$, evolve on super-Hubble scales until some time in matter (or
radiation) era, when they cross the Hubble scale for the second time at $q=a_{\rm 2x}H_{\rm 2x}$.
At the end of inflation the ratio of the physical wave length to the Hubble scale will be,
$(a/k)/H^{-1}\propto a^{1-\epsilon}$, where we assumed that $H\propto a^{-\epsilon}$ and
$\epsilon\ll 1$ and constant. Since during radiation and matter era, $H^{-1}\propto a^2$ and
$H^{-1}\propto a^{3/2}$, the following relation holds,
\begin{equation}
  \bigg(\frac{a_{\rm end}}{a_{\rm 1x}}\bigg)^{1-\epsilon}
  =\frac{a_{\rm eq}}{a_{\rm end}}\times
  \bigg(\frac{a_{\rm 2x}}{a_{\rm eq}}\bigg)^{1/2}
\,, \label{ratio of scale factors}
\end{equation}
where $a_{\rm eq}$ is the scale factor at radiation-matter equality. Neglecting $\epsilon\ll 1$ and
taking account of the definition, $N_{\rm 1x} = \log(a_{\rm end}/a_{\rm 1x})$, Eq.~(\ref{ratio of
scale factors}) yields
\begin{equation}
  N_{\rm 1x} \approx \log\bigg(\frac{a_{\rm eq}}{a_{\rm end}}\bigg)
      + \frac12\log\bigg(\frac{a_{\rm 2x}}{a_{\rm eq}}\bigg)
\,.\label{Crossings}
\end{equation}
Now, recalling that $a_0/a_{\rm eq}\sim 3200$, and $\log(a_{\rm eq}/a_{\rm end})=(1/2)\log(H_{\rm
end}/H_{\rm eq})$, $H_0=1.5\times 10^{-42}~{\rm GeV}$, $H_{\rm eq}\simeq 3\times 10^{-37}~{\rm
GeV}$, Eq.~(\ref{Crossings}) can be recast as
\begin{equation}
N_{\rm 1x} \simeq %\frac12[50\log(10)-\log(3)+\log(3200)]
61+\frac{1}{2}\log\Big(\frac{H_{\rm end}}{10^{13}~{\rm GeV}}\Big) -
\frac12\log(1+z_{\rm 2x}), \label{Crossings:2}
\end{equation}
where $z_{\rm 2x}=(a_0/a_{\rm 2x})-1$ is the redshift at the second Hubble crossing and $a_0$ is
the scale factor today, such that the last term in~(\ref{Crossings:2}) drops out for modes of the
Hubble length today~\footnote{The absence of the graviton signal in the CMB limits $H_{\rm end}\leq
3\times 10^{13}~{\rm GeV}$ ($V_{\rm end}^{1/4}<1.1\times 10^{16}~{\rm GeV}$), and hence $N_{\rm
1x}\simeq 61.6$ is an upper limit for the modes with $z_{\rm 2x}\simeq 0$. In Fig.~\ref{fig:
HorizonCrossing} $N_{\rm 1x}=60$ e-folds corresponds to about $H_{\rm end}\simeq 10^{12}~{\rm GeV}$
($V_{\rm end}^{1/4}\simeq 2\times 10^{15}~{\rm GeV}$).}.
\paragraph{}
The black holes that we have so far considered formed during a pre-inflationary era ($N > 60$).
From Fig.~\ref{fig: HorizonCrossing} we see that if a black hole forms between $N=70$ and $N=60$
and is within our past light-cone then the scales $q/H_0 \sim 1-100$ will correspond to observable
scales today. Moreover, their amplitude on the CMB will be of the order of $10\%$ of the signal
obtained for a homogeneous background, as can be seen
from our numerical results in Figs.~\ref{fig:
qy1Spectrum}--\ref{fig: SpectrumIsotropic}. Such a strong signal has not been observed, implying
that such black holes had not formed. If inflation lasted much longer than 60 e-foldings, it would
be useful to estimate the probability for black hole formation during inflation. In order to
estimate the formation probability in the early stages of inflation, we first note that on
sub-Hubble scales Eq.~\eqref{DensityPerturbEvolution} for density perturbations can be used in
inflation~\footnote{{\it cf.} also Lyth and Liddle `The primordial density perturbation', Cambridge
University Press (2009)} with the equation of state parameter $w\approx 0$ for matter and the
Hubble rate $H \simeq H_0 a^{-\epsilon}$. In slow-roll inflation we can take the Hubble rate to be
constant to estimate the growth of the perturbations. Since $\rho_0(t) \sim 1/a^{3(1+w)}$, the last
term in~\eqref{DensityPerturbEvolution} becomes rapidly negligible and we find that the evolution
of the perturbation is given by
\begin{equation} \label{evolutionDeltaInflation}
\delta(k,t) = \delta_*(k) + \overline{\delta}_*(k){\rm e}^{-2\int H dt}\,,
\end{equation}
which means that the leading perturbation tends to a constant. Using the relation between the
density perturbation and the fluctuation in the number of particles in the comoving ball with
physical radius $R$, Eq.~\eqref{delta N over N}, it follows that during inflation ($t>t_I$)
\begin{equation}
\frac{\delta N(R,t)}{\langle N(R,t)\rangle} \approx \frac{\delta N(R,t_I)}{\langle N(R,t_I)\rangle}
\end{equation}
where $t_I$ denotes the beginning of inflation. Therefore, also the variance in the fluctuations of
$N(R,t)$ is constant for $t>t_I$,
\begin{equation}
\sigma (R,t) = \langle \delta N^2(R,t)\rangle = \frac{\langle \delta N^2(R,t_I) \rangle}{\langle
N(R,t_I)\rangle^2}\langle N(R,t)\rangle^2 = \langle \delta N^2(R,t_I) \rangle
\end{equation}
because the average number of particles in a comoving ball does not change during
inflation~\footnote{To see this consider the geodesic equation for particles during the
inflationary phase, $\dot{\vec v} + H \vec v \approx 0$, which shows that their motion (with
respect to the fluid rest frame) is exponentially damped, $\vec v (t) = \vec v (t_I) {\rm
e}^{-(N_I-N(t))}$. This means that very quickly particles get frozen and cannot leave the comoving
volume, leading to a constant particle number per comoving volume.}. Since the universe is
dominated by the potential energy of the inflaton field, Eq.~\eqref{average:N} is not valid during
inflation. Taking this into account, the Friedmann equation is
\begin{equation}
H^2 = \frac{8\pi}{3m_P^2}\left(\rho_0(t)+ V(\phi)\right) \,,
\end{equation}
and we find that the critical fluctuation that is necessary for black hole formation is given by
$\delta N_{\rm cr}(R,t) = m_p^2R/(2m) - \langle N(R,t)\rangle \approx m_p^2R/(2m)$. Considering
fluctuations in a volume of radius $R$, we thus obtain the result
\begin{equation} \label{ExpSupprInflation}
\frac{\delta N_{\rm cr}(R,t)}{\sqrt{2\sigma(R,t)}} = \frac{m_P^2 R}{2m\sqrt{2\langle \delta
N^2(R,t_I)\rangle}}\,.
\end{equation}
In the case when $R$ is the comoving radius, $R=R_I {\rm e}^{N_I-N(t)}$, the expression
in~\eqref{ExpSupprInflation} grows exponentially in physical time. On the other hand, when $R$ is a
constant physical radius, yields $\delta N_{\rm cr}/\sqrt{2\sigma} \varpropto {\rm
e}^{\frac{3}{2}(N_I-N(t))}$. We conclude from Eq.~\eqref{probability:BH formation} that the
probability for black hole formation is rapidly (faster than exponentially) suppressed during
inflation. That means that if inflation lasts much longer than about 70 e-foldings the probability
of observing any inflationary black holes within our past light-cone will be tiny. Furthermore,
from Figs.~\ref{fig: Factorisation} and~\ref{fig: HorizonCrossing} we see that the black holes
which form later during inflation (corresponding to $N\leq 50$) are observed as $q\ll H_0$ and
induce unobservably small fluctuations in the CMB. According to the above analysis the probability
for formation of these black holes is tiny.

 We shall now comment on various ways how this work can be extended. The case of a black hole in
inflation can be easily generalized to heavy particles and in particular monopoles by gluing the
exterior Schwarzschild-de~Sitter geometry to an interior matter solution. The possibility of
magnetically charged black holes in the presence of magnetic monopoles is discussed in
\cite{Hiscock:1983ag}. Also, it would be interesting to study a rotating body giving rise to an
exterior Kerr-de~Sitter geometry. Furthermore, there is no reason why a black hole should form in
the rest frame of the inflaton field. Assuming a black hole forms with an initial relative speed
$v_0$, its motion will be damped the same way as the motion of a test particle, which obeys a
geodesic equation, $\dot{\vec v} + H \vec v \approx 0$, such that the speed is Hubble damped as
$\vec v = {\vec v}_0/a$ after the formation. Nevertheless, in the early stages after formation the
speed of the black hole can yield significant velocity perturbations which may have observational
impact on the structure formation, and an interesting question is whether a relation can be
established with the recent observation of large scale flows~\cite{Kashlinsky:2009dw}. It is also
worth investigating what is the damping of the speed of a spinning massive object.

 When this work was nearing completion, we became aware of
Refs.~\cite{Cho:2009en,Aguilar:2010jk,Reyes:2010tq} which, just like this work, address the
problem of scalar field fluctuations on a Schwarz-schild-de Sitter (SdS) background. We note however
that there are important differences between these papers and ours, and that our analysis of the
black hole perturbations on SdS background is much more elaborate~\footnote{The most important
difference is that, in contrast to this work, Cho, Ng and Wang~\cite{Cho:2009en} do not make any
attempt to relate the inflaton fluctuation to the comoving curvature perturbation, which can be
defined on the Schwarzschild-quasi de Sitter space, and which is the correct generalization of
homogeneous inflationary spaces. Furthermore, we provide an analytic estimate for the formation
probability of small pre-inflationary and inflationary black holes. Next, we provide an exact
answer for the Schwinger-Keldysh propagators in the double momentum space (whereby admittedly
treating the black hole as a small perturbation on de Sitter space), and we provide a more complete
analysis of the significance of the results by discussing how the curvature spectrum gets perturbed
as a function of the position of the black hole.}. Since the first version of the present paper,
several articles have appeared which discuss anomalies in the CMB from black holes and point-like
defects. In a recent proposal by Gurzadyan and Penrose~\cite{Gurzadyan:2010da,Gurzadyan:2010xj} the
attempt was made to relate the presence of (anomalous) concentric circles in the CMB sky to pre-Big
Bang activities involving black holes in the framework of conformal cyclic cosmology. This has been
the subject of recent debates~\cite{Wehus:2010pj,Hajian:2010cy}. The role of the presence of
pre-inflationary particles for anomalies in the CMB was investigated in~\cite{Fialkov:2009xm}
and~\cite{Kovetz:2010kv} where the imprint is also expected to be given in form of rings. Finally,
the effect of pre-inflationary black holes on the CMB power spectrum was studied
in~\cite{Scardigli:2010gm} and a link to the quadrupole anomaly was made.

\vspace{.7cm}

\noindent {\bf Acknowledgements.} We thank Igor~Khavkine, Renate~Loll and Albert~Roura for helpful
discussions.

\section*{Appendix A} \label{AppendixA}

The coordinate transformations that give rise to the line element in cosmological form,
Eq.~(\ref{cosmologicalSdS}), are presented here.
\paragraph{}
In its static form the Schwarzschild de Sitter (SdS) solution is given by the line element
\begin{eqnarray}
ds^2 &=& -f(\tilde{r})dt^2+ f(\tilde{r})^{-1}d\tilde{r}^2 + \tilde{r}^2d\Omega^2 \\
f(\tilde{r}) &=& 1-2GM/\tilde{r}- \tilde{r}^2/R_0^2\,, \nonumber
\end{eqnarray}
with $R_0 = \sqrt{3/\Lambda}$ being the Hubble radius and $\Lambda$ the cosmological constant. This
line element reflects the spherical symmetry and the time translation symmetry.

 If we take
\begin{equation}
\tau(t,\tilde{r}) = t - \int {\sqrt{1-f(\tilde{r})}\over f(\tilde{r})}d\tilde{r}, \quad R(\tau,\tilde{r})=\tilde{r}/a(\tau)
\end{equation}
for $a(\tau)= {\rm e}^{\tau/R_0}$ then the metric becomes
\begin{equation}
ds^2 = -d\tau^2 + a^2(\tau)\left[(dR + F(\tau,R)d\tau)^2 + R^2d\Omega^2 \right]
\end{equation}
with
\begin{equation}
F(\tau,R) = \frac{R}{R_0} \left(1 - \sqrt{{2GM R_0^2 \over R^3 a^3(\tau)} + 1}\right).
\end{equation}
The spatial slices $\tau = {\rm const.}$ are flat. In order to make the metric diagonal, we have to
solve a differential equation for a function $R(\tau,r)$
\begin{equation}
\frac{\partial R}{\partial \tau} = -F(\tau,R),
\end{equation}
which has the solution
\begin{equation}
R(\tau,r) = (2GMR_0^2)^{1/3}{\rm e}^{-\tau/R_0}\sinh^{2/3}\left(\frac{3}{2}\frac{\tau+\tau_0(r)}{R_0}\right),
\end{equation}
where $\tau_0(r)$ is a $\tau$-independent integration constant. This constant can be determined by
the requirement that $R(\tau,r) = r$ for $M=0$, \textit{i.e.}\ homogeneous cosmology is recovered.
One finds that
\begin{equation}
\tau_0(r)=R_0\log \left(\left(\frac{2}{GMR_0^2}\right)^{1/3}r\right),
\end{equation}
and hence,
\begin{equation}
R(\tau,r) = r\left(1-\frac{GMR_0^2}{2a^3r^3}\right)^{2/3}. \nonumber
\end{equation}
The metric then becomes
\begin{align}\label{CosmologicalMetric}
ds^2 &= -d\tau^2 + a^2(\tau)\left[\left({\partial R \over \partial r} (\tau,r) \right)^2 dr^2 + R^2 (\tau,r)d\Omega^2 \right] \\
&= -d\tau^2 + \frac{(2GMR_0^2)^{2/3}}{r^2}\sinh^{4/3}\left(\frac{3}{2}\frac{\tau+\tau_0(r)}{R_0}\right)\left[\coth^2 \left(\frac{3}{2}\frac{\tau+\tau_0(r)}{R_0}\right)dr^2 + r^2d\Omega^2 \right] \nonumber \\
&= -d\tau^2 + a^2(\tau)\left(1-\frac{GMR_0^2}{2a(\tau)^3r^3}\right)^{4/3}\left[\left(\frac{1+\frac{GMR_0^2}{2a(\tau)^3r^3}}{1-\frac{GMR_0^2}{2a(\tau)^3r^3}}\right)^2dr^2 + r^2d\Omega^2 \right], \nonumber
\end{align}
which is recast in the main text, Eq.~(\ref{cosmologicalSdS}), in conformal time $\eta$,
$d\eta=dt/a(t)$. Note that the metric is singular at
\begin{equation}
r_0(\tau) = \left(\frac{GMR_0^2}{2}\right)^{1/3}\frac{1}{a(\tau)},
\end{equation}
and regular for all values $r>r_0(\tau)$. As a check, $R_{\mu\nu}-\Lambda g_{\mu\nu}=0$, and the
Kretschmann invariant is
\begin{equation}
R_{\mu\nu\rho\sigma}R^{\mu\nu\rho\sigma}
= \frac{48GM^2}{\tilde{r}^6}+\frac{8\Lambda^2}{3}
= \frac{48GM^2}{a(\tau)^6r^6\left(1-\frac{3GM}{2\Lambda a(\tau)^3r^3}\right)^4}
  + \frac{8\Lambda^2}{3}
\,,
\label{Riemann squared}
\end{equation}
with $\tilde{r}$ denoting as previously the Schwarzschild radial coordinate. This proves that $r_0$
corresponds to the curvature singularity at $\tilde{r}=0$. Since $\partial \tilde{r}/\partial r>0$
for $r>r_0(\tau)$, we conclude that $\tilde{r} \to \infty$ when $r \to \infty$ without any
double-valued regions.

\section*{Appendix B}

For the derivation of~(\ref{CorrectedGreensFunctionMomentumSpace}) from
(\ref{CorrectedGreensFunctionPositionSpace}) we have to evaluate first
\begin{align}
I_1 (\vec{k},\vec{k}',\eta'') &\equiv \int_{-\mu \eta''}^\infty \frac{dr''}{r''} \int_{-1}^1 d\cos\theta''\int_0^{2\pi} d\phi'' {\rm e}^{i(\vec{k}'-\vec{k})\cdot\vec{x}''} = 4\pi\left(\frac{\sin\rho}{\rho}-\mathrm{Ci}(\rho)\right) \\
I_2 (\vec{k},\vec{k}',\eta'') &\equiv \int_{-\mu \eta''}^\infty \frac{dr''}{r''} \int_{-1}^1 d\cos\theta''\int_0^{2\pi} d\phi'' \frac{(\vec{k}'\cdot\vec{x}'')^2}{r''^2} {\rm e}^{i(\vec{k}'-\vec{k})\cdot\vec{x}''} \nonumber \\
&= \frac{4\pi k'^2}{3} \left(\left(\frac{\sin\rho}{\rho}-\mathrm{Ci}(\rho)\right) + \left(3\cos^2\widetilde{\theta} -1\right)\left(\frac{\cos\rho}{\rho^2}-\frac{\sin\rho}{\rho^3}\right)\right), \nonumber
\end{align}
with the cosine integral function $\mathrm{Ci}(z)=-\int_z^\infty
dt \cos(t)/t$, $\widetilde{\theta} =
\sphericalangle(\vec{k}'-\vec{k},\vec{k}')$ and
$\rho=-\mu\|\vec{k}-\vec{k}'\|\eta''$. Then, it follows that
\begin{align}
I(\vec{k},\vec{k}',\eta'') &\equiv \int_{-\mu\eta''}^\infty \frac{dr''}{r''} \int_{-1}^1d\cos\theta''\int_0^{2\pi}d\phi''\left(-k'^2+3\frac{(\vec{k}'\cdot\vec{x}'')^2}{r''^2} \right) {\rm e}^{i(\vec{k}'-\vec{k})\cdot\vec{x}''} \\
&= -k'^2I_1(\vec{k},\vec{k}',\eta'') + 3I_2(\vec{k},\vec{k}',\eta'') \nonumber \\
&= 4\pi
k'^2\left(3\cos^2\widetilde{\theta}-1\right)\left(\frac{\cos(\mu\|\vec{k}-\vec{k}'\|\eta'')}{\mu^2\|\vec{k}-\vec{k}'\|^2\eta''^2}-\frac{\sin(\mu\|\vec{k}-\vec{k}'\|\eta'')}{\mu^3\|\vec{k}-\vec{k}'\|^3\eta''^3}\right).
\nonumber
\end{align}
To establish~(\ref{FeynmanPropagators}) and~(\ref{FeynmanPropagatorsAnti}) we used the following
relations for the step function:
\begin{align}
&\Theta(\eta-\eta'')\Theta(\eta''-\eta') = \Theta(\eta-\eta')\left[\Theta(\eta''-\eta')-\Theta(\eta''-\eta) \right]
\label{Theta functions:relations}
\\
&\Theta(\eta-\eta'')\Theta(\eta'-\eta'') = \Theta(\eta-\eta')\Theta(\eta'-\eta'')+\Theta(\eta'-\eta)\Theta(\eta-\eta'') \nonumber \\
&\Theta(\eta''-\eta)\Theta(\eta''-\eta') = \Theta(\eta-\eta')\Theta(\eta''-\eta)+\Theta(\eta'-\eta)\Theta(\eta''-\eta') \nonumber \\
&\Theta(\eta''-\eta)\Theta(\eta'-\eta'') = \Theta(\eta'-\eta)\left[\Theta(\eta''-\eta)-\Theta(\eta''-\eta') \right]. \nonumber
\end{align}
The final form for $J_{+-,+-}$ and $J_{+-,-+}$, Eqs.~(\ref{FinalJs}) and~(\ref{FinalJsB}), can be
obtained by solving the following integral, for real parameters $A,B,\alpha$ and $\alpha\neq 1$ and
$\rho = -\mu\eta''p$ with $A+\alpha B=0$,
\begin{align}
&\int d\rho \left(\rho^3 + iA\rho^2 + B\rho\right)\left(\frac{\cos\rho}{\rho^2}-\frac{\sin\rho}{\rho^3}\right) {\rm e}^{i\alpha\rho} \\
&= \left[\left(\frac{2}{(\alpha^2-1)^2}+\frac{\alpha (A-i\rho)}{\alpha^2-1}\right)\cos\rho + \left(\frac{i(\alpha-A)-\rho}{\alpha^2-1}-\frac{2i\alpha}{(\alpha^2-1)^2}+\frac{B}{\rho}\right)\sin\rho\right]e^{i\alpha\rho}. \nonumber
\end{align}

\section*{Appendix C}
\label{AppendixC}

We present here the general integral expression for the finite part of the spectrum. For this we
write
\begin{align}
\vec{q} &= q(0,0,1) \\
\vec{p} &= p(\cos\phi\sin\theta,\sin\phi\sin\theta,\cos\theta) \nonumber \\
\vec{y} &= y(\cos\phi_y\sin\theta_y,\sin\phi_y\sin\theta_y,\cos\theta_y) \nonumber
\end{align}
Introducing $x=\cos\theta$ and $w=p/(2q)$, as in the main text, we find that
\begin{equation}
\int_0^{2\pi}d\phi\cos(\vec{p}\cdot\vec{y}) = 2\pi J_0 (2qyw\sqrt{1-x^2}\sin\theta_y)\cos(2qywx\cos\theta_y).
\end{equation}
In terms of the variables $\kappa$ and $\kappa'$,
\begin{align}
\kappa = k/q = \sqrt{1+2wx+w^2}, \qquad \kappa' = k'/q = \sqrt{1-2wx+w^2},
\end{align}
the integral form of the spectrum, in Eqs.~(\ref{PropagatorSplitIntoFiniteAndIRpart}) and
(\ref{PropagatorToSpectrum}--\ref{SplitSpectrumIntoFiniteAndIRpart}), becomes
\begin{align} \label{CorrectionFullIntegralExpression}
&\delta\mathcal{P}^{\mathrm{fin}}(q,y,\eta) = \frac{2\mu q H_0}{3\pi^3} \int_0^\infty dw \int_{-1}^1 \frac{dx}{\kappa^2}\left(2+\frac{3(x^2-1)}{\kappa'^2}\right) \\
&\times \Bigg\{\frac{\cos((\kappa+\kappa')q/H_0)}{\kappa+\kappa'}\left(\cos(2\mu wq/H_0)-\frac{\sin(2\mu wq/H_0)}{2\mu wq/H_0}\right) \nonumber \\
&\quad + (2\mu w)^2 \bigg[\frac{\cos((\kappa+\kappa')q/H_0)}{(\kappa+\kappa')^3}\left(\cos(2\mu wq/H_0)+\frac{\sin(2\mu wq/H_0)}{2\mu wq/H_0}\right) \nonumber \\
&\quad - \frac{2\sin((\kappa+\kappa')q/H_0)}{(q/H_0)(\kappa+\kappa')^4}\cos(2\mu wq/H_0) + \frac{(q/H_0) \sin((\kappa+\kappa')q/H_0)}{(\kappa+\kappa')^2}\frac{\sin(2\mu wq/H_0)}{2\mu wq/H_0}\bigg] \Bigg\} \nonumber \\
&\times J_0 (2qyw\sqrt{1-x^2}\sin\theta_y)\cos(2qywx\cos\theta_y) \nonumber \\
&\qquad\qquad\qquad -\frac{2\mu q H_0}{3\pi^3} \int_0^\infty dw \int_{-1}^1 \frac{dx}{\kappa^{3}}\Bigg[\left(2+\frac{3(x^2-1)}{\kappa'^2}\right) \nonumber \\
&\times \Bigg\{\frac{\sin((\kappa+\kappa')q/H_0)}{(q/H_0)\kappa'}\left(\cos(2\mu wq/H_0)-\frac{\sin(2\mu wq/H_0)}{2\mu wq/H_0}\right) \nonumber \\
&\quad -(2\mu w)^2 \bigg[\frac{\cos((\kappa+\kappa')q/H_0)}{\kappa'(\kappa+\kappa')}\frac{\sin(2\mu wq/H_0)}{2\mu wq/H_0} - \frac{\sin((\kappa+\kappa')q/H_0)}{(q/H_0)\kappa'(\kappa+\kappa')^2}\cos(2\mu wq/H_0)\bigg]\Bigg\} \nonumber \\
&\times J_0 (2qyw\sqrt{1-x^2}\sin\theta_y)\cos(2qywx\cos\theta_y) \nonumber \\
&- 2\Bigg\{\frac{\sin(2q/H_0)}{2q/H_0}\left(\cos(2\mu q/H_0)-\frac{\sin(2\mu q/H_0)}{2\mu q/H_0}\right) \nonumber \\
&\quad - \mu^2 \bigg[\cos(2q/H_0)\frac{\sin(2\mu q/H_0)}{2\mu q/H_0} - \frac{\sin(2q/H_0)}{2q/H_0}\cos(2\mu q/H_0)\bigg]\Bigg\}\cos(2qy\cos\theta_y)\Bigg]. \nonumber
\end{align}
For $q/H_0 \ll 1$ one finds
\begin{equation}
\delta\mathcal{P}_\phi(q,y,\eta) = \frac{16 \mu^5 q^3}{9\pi^3H_0}F(qy, \theta_y) + \frac{16 \mu^5 q^3}{9\pi^3H_0}\left(\log\left(\frac{q^2}{k_0^2}\right)+2\right)\cos(2qy\cos\theta_y)
\end{equation}
with
\begin{align} \label{FactorizationFunctionF}
&F(qy, \theta_y) = \\
&+ \int_0^\infty dw\ w^4 \int_{-1}^1 \frac{dx}{\kappa^2 (\kappa+\kappa')^3} \nonumber \\
&\qquad\qquad\qquad\qquad \times \left(2+\frac{3(x^2-1)}{\kappa'^2}\right) 2J_0 (2qyw\sqrt{1-x^2}\sin\theta_y)\cos(2qywx\cos\theta_y) \nonumber \\
& + \int_0^\infty dw\ \int_{-1}^1 \frac{dx}{\kappa^{3}}\Bigg[\left(2+\frac{3(x^2-1)}{\kappa'^2}\right) \frac{2w^4J_0 (2qyw\sqrt{1-x^2}\sin\theta_y)\cos(2qywx\cos\theta_y)}{\kappa'(\kappa+\kappa')} \nonumber \\
&\qquad\qquad\qquad\qquad\qquad -\cos(2qy\cos\theta_y)\Bigg], \nonumber
\end{align}

\section*{Appendix D}
\label{AppendixD}

 This appendix contains the fitting functions for the numerical data. In the isotropic case
$\vec{y}=0$, the correction to the spectrum normalized by the scale invariant de~Sitter spectrum
$4\pi^2/H_0^2$ has been fitted in Fig.~\ref{fig: SpectrumIsotropic} with the following function:
\begin{align}
&f(q,\mu,k_0) = \mu A(\mu) \log(q/H_0)\Big[(1-\cos(2\mu q/H_0))\sin(2q/H_0-1/4) \\
&\qquad\qquad + \mu^2 B(\mu) (q/H_0) \cos(2q/H_0-\pi/4)\Big] \nonumber \\
&\quad + \frac{16 \mu q/H_0}{3\pi} \Bigg\{\frac{\sin(2q/H_0-1/4)}{2q/H_0}\left(\cos(2\mu q/H_0)-\frac{\sin(2\mu q/H_0)}{2\mu q/H_0}\right) \nonumber \\
&\qquad - \mu^2 \bigg[\cos(2q/H_0-1/4)\frac{\sin(2\mu q/H_0)}{2\mu q/H_0} - \frac{\sin(2q/H_0-1/4)}{2q/H_0}\cos(2\mu q/H_0)\bigg]\Bigg\} \nonumber \\
&\qquad \times \Big(\log(3q^2)+2\Big) \nonumber \\
&\quad - \frac{16 \mu q/H_0}{3\pi} \Bigg\{\frac{\sin(2q/H_0)}{2q/H_0}\left(\cos(2\mu q/H_0)-\frac{\sin(2\mu q/H_0)}{2\mu q/H_0}\right) \nonumber \\
&\qquad - \mu^2 \bigg[\cos(2q/H_0)\frac{\sin(2\mu q/H_0)}{2\mu q/H_0} - \frac{\sin(2q/H_0)}{2q/H_0}\cos(2\mu q/H_0)\bigg]\Bigg\}\left(\log\left(\frac{q^2}{k_0^2}\right)+2\right) \nonumber.
\end{align}
We have no analytic expression for the functions $A$ and $B$ but we observe that the dependence on
$\mu$ is weak,
\begin{align}
&A(1/50) = 0.76,\qquad B(1/50) = -1.0. \\
&A(1/20) = 0.87,\qquad B(1/20) = -0.7, \nonumber \\
&A(1/10) = 0.80,\qquad B(1/10) = -1.3, \nonumber
\end{align}
Obviously, setting $A=0.8$ and $B=-1.0$ for all $\mu$ results in reasonable fits, too.

\end{document}